\RequirePackage{fix-cm}
\documentclass{svjour3}                     % onecolumn (standard format)
\usepackage{amsmath,amsfonts,amssymb,hyperref,lscape,float,pdflscape,setspace,flexisym,color,physics}
\usepackage{lineno,graphicx,multirow,hhline,nicefrac,amsmath,rotating,epstopdf,mathrsfs,nameref}
\usepackage[round,sort,comma,numbers]{natbib}
\smartqed  % flush right qed marks, e.g. at end of proof

\makeatletter
\def\@biblabel#1{}
\makeatother

\begin{document}
	
\title{A continuum mathematical model of substrate-mediated tissue growth}
	
\author{Maud El-Hachem$^{1}$  \and
		Scott W McCue$^{1}$	\and
		Matthew J Simpson$^{1*}$
}
	
\institute{$^{1}$ School of Mathematical Sciences, Queensland University of Technology, Brisbane, Australia \\
		\email{matthew.simpson@qut.edu.au}
}
	
\date{Received: date / Accepted: date}

\maketitle

\begin{abstract}
We consider a continuum mathematical model of biological tissue formation inspired by recent experiments describing thin tissue growth in 3D--printed bioscaffolds.  The continuum model involves a partial differential equation describing the density of tissue, $\hat{u}(\hat{\mathbf{x}},\hat{t})$, that is coupled to the concentration of an immobile extracellular substrate, $\hat{s}(\hat{\mathbf{x}},\hat{t})$.  Cell migration is modelled with a nonlinear diffusion term, where the diffusive flux is proportional to $\hat{s}$, while a logistic growth term models cell proliferation.  The extracellular substrate $\hat{s}$ is produced by cells, and undergoes linear decay.  Preliminary numerical simulations show that this mathematical model, which we call the \textit{substrate model}, is able to recapitulate key features of recent tissue growth experiments, including the formation of sharp fronts.  To provide a deeper understanding of the model we then analyse travelling wave solutions of the substrate model, showing that the model supports both sharp--fronted travelling wave solutions that move with a minimum wave speed, $c = c_{\rm{min}}$, as well as smooth--fronted travelling wave solutions that move with a faster travelling wave speed, $c > c_{\rm{min}}$.  We provide a geometric interpretation that explains the difference between smooth-- and sharp--fronted travelling wave solutions that is based on a slow manifold reduction of the desingularised three--dimensional phase space.  In addition to exploring the nature of the smooth-- and sharp--fronted travelling waves, we also develop and test a series of useful approximations that describe the shape of the travelling wave solutions in various limits.  These approximations apply to both the sharp--fronted travelling wave solutions, and the smooth--fronted travelling wave solutions.  Software to implement all calculations is available at \href{https://github.com/ProfMJSimpson/Substrate_Mediated_Invasion}{GitHub}.

\keywords{Travelling Wave, Tissue Engineering, Partial differential equation, Fisher-KPP, Porous-Fisher, Diffusion, Logistic growth.}
\end{abstract}

\begin{spacing}{2.00}

\newpage
\section{Introduction}
\begin{landscape}
	\begin{figure}[h!]
		\centering
		\includegraphics[width=1\linewidth]{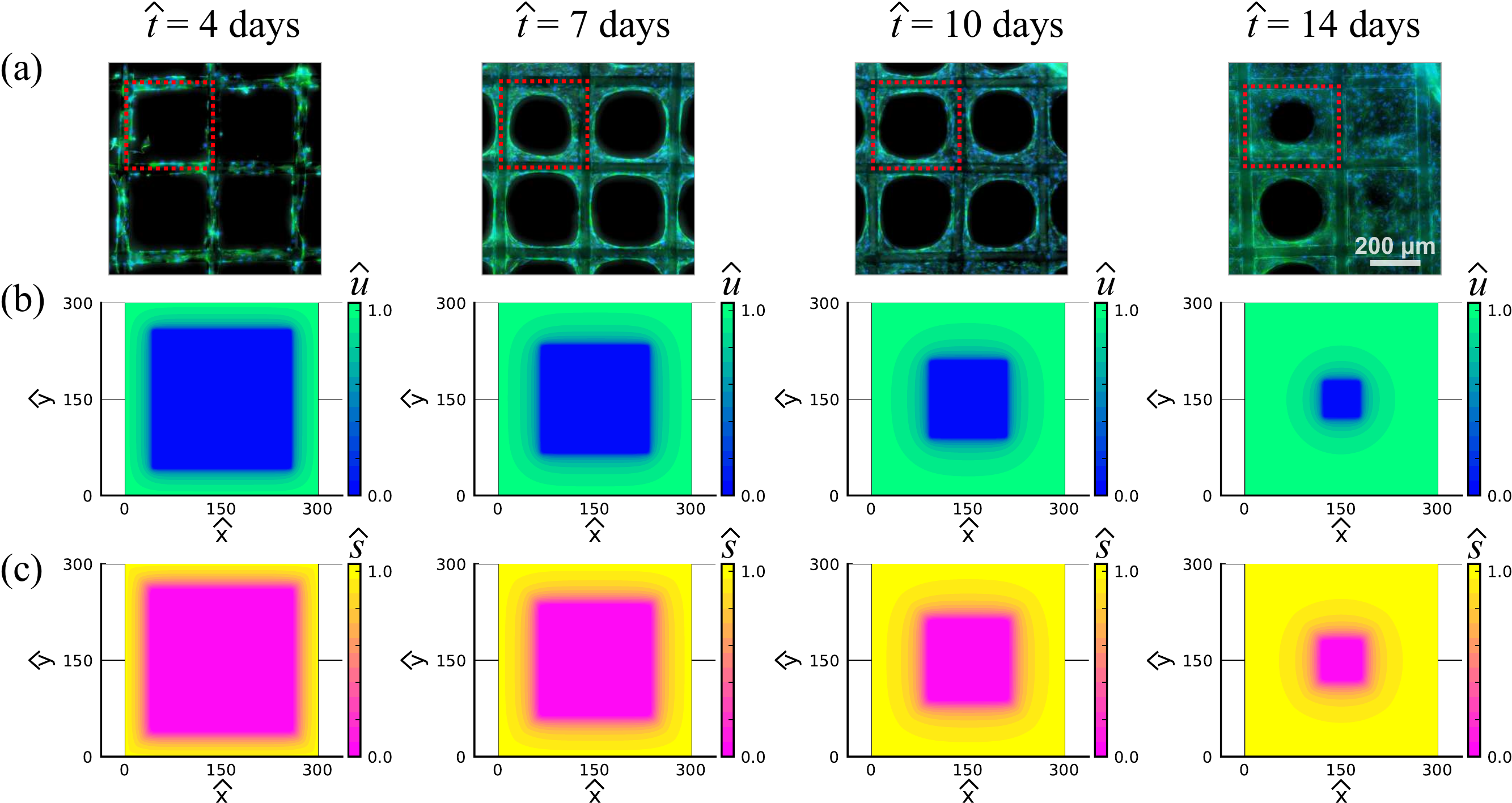}
		\caption{\textbf{Experimental and simulated osteoblast tissue formation within a square--shaped 3D--printed pore.} (a) Composite fluorescence microscopy images of pore bridging experiments~\cite{Buenzli2020,Browning2021}. Cell nuclei are shown in blue,  tissue and cytoskeleton are shown in green.  Each subfigure shows four adjacent square pores, each with side length of $\hat{L}=300$ $\mu$m, and images are shown at various times, $\hat{t} = 4, 7, 10$ and $14$ days, as indicated.  For clarity,  in each subfigure we outline the border of the upper--left pore (red dashed).    Experimental images are reproduced from~\cite{Buenzli2020} with permission. (b)--(c)  Numerical solution of Equations (\ref{eq:GouvdiffUDimensional2D})--(\ref{eq:GouvdiffSDimensional2D}) on a square domain with side length $\hat{L}=300$ $\mu$m.  (b) Evolution of $\hat{u}$. (c) Evolution of $\hat{s}$.  Each column of the figure corresponds to $\hat{t} = 4, 7, 10$ and $14$ days, as indicated.  Parameter values for the mathematical model are $\hat{D}=300$~$\mu$m$^2$/day, $\hat{\lambda} = 0.6$ /day, $\hat{K}_u = 1$ cells/$\mu$m$^2$, $\hat{K}_s = 1$ mol/$\mu$m$^2$, $\hat{r}_1 = 1$ mol/(cells day),  $\hat{r}_2 = 1$ /day.  The numerical solution of (\ref{eq:GouvdiffUDimensional2D})--(\ref{eq:GouvdiffSDimensional2D}) is obtained on a $101\times101$ mesh, and temporal integration is performed with uniform time steps of duration $\Delta \hat{t} = 1\times10^{-2}$ day.}
		\label{fig:1}
	\end{figure}
\end{landscape}

Over the last decade, tissue engineering has been revolutionised through the use of 3D printing technologies that produce 3D bioscaffolds upon which $\textit{in vitro}$ tissues can be grown in biologically realistic geometries~\cite{Ambrosi2019,Dzobo2018}. \textit{In vitro} tissues grown on 3D scaffolds are more reproducible and more biologically realistic than tissues grown in traditional two--dimensional tissue culture~\cite{Lanaro2021}.  The experimental images in Figure \ref{fig:1}(a) show the evolution of thin 3D tissues that are produced by seeding a 3D--printed scaffold with osteoblast precursor cells~\cite{Buenzli2020,Browning2021}.  In this experiment, cells are seeded onto the perimeter of 3D--printed square shaped pores, where each pore has sides of approximately $300$ $\mu$m in length.  Each subfigure in Figure \ref{fig:1}(a) shows four adjacent pores.  As the experiment proceeds, individual cells migrate off the scaffold into the pore, and then combined cell migration and cell proliferation leads to the formation of a sharp-fronted tissue profile that invades into the pore.  This process eventually forms a thin tissue that closes or \textit{bridges} the pore after approximately 14 days~\cite{Buenzli2020,Browning2021}.  A notable feature of these experiments is the fact that tissue formation involves a well--defined moving front that is very obvious in Figure \ref{fig:1}(a).  Closer inspection of these experimental images shows that cells not only migrate and proliferate during the pore bridging process, but cells also produce an extracellular medium that is laid down onto the surface of the pore~\cite{Lanaro2021}.

Continuum mathematical models of tissue formation have a long history, with many early models based on the classical Fisher--KPP model~\cite{Ablowitz11979,Canosa1973,Fisher1937,Kolmogorov1937}.  The Fisher--KPP model describes cell migration using a one dimensional linear diffusion term, and cell proliferation is modelled using a logistic source term.  Many different types tissue formation experiments have been successfully modelled using the Fisher--KPP model~\cite{Maini2004,Jin2016,Johston2015,Warne2019} or two-dimensional extensions of the Fisher--KPP model~\cite{Sherratt1990,Simpson2013,Swanson2003}.  While these studies show that simple mathematical models based on the Fisher-KPP framework successfully capture certain features of tissue formation, there are several well--known limitations of the Fisher--KPP model that can be addressed by considering extensions of that model~\cite{Murray2002}.  One such criticism is that the linear diffusion term in the Fisher--KPP model leads to smooth density profiles that do not represent well--defined fronts, such as those we see in Figure \ref{fig:1}(a).

One way to overcome this limitation is to work with the Porous--Fisher model where the linear diffusion term is generalised to a degenerate nonlinear diffusion term with a power law diffusivity~\cite{Fadai2020,Sanchez1994,Sengers2007,Witelski1994,Witelski1995}.  While the Porous--Fisher model leads to sharp--fronted density profiles, this approach introduces a separate complication of having to justify the choice of the exponent in the power law diffusivity~\cite{Jin2016,McCue2019,Sherratt1990,Simpson2011,Warne2019}.  A further weakness of both the Fisher--KPP and Porous--Fisher models is that they deal with a single species, such as a density of cells, and do not explicitly describe how the population of cells invades into surrounding cells, or interacts with the surrounding environment.  This second limitation has been addressed by introducing more complicated mathematical models, such as the celebrated Gatenby--Gawlinski model of tumour invasion~\cite{Gatenby1996}, that explicitly describes how a population of tumour cells degrades and invades into a population of surrounding healthy tissue by explicitly modelling both populations and their interactions.  Since the Gatenby--Gawlinski framework was proposed in 1996, subsequent studies have since analyzed the relationship between individual--level mechanisms and the resulting population--level continuum descriptions~\cite{Painter2003}, calibrating these mathematical models to match experimental measurements of melanoma invasion~\cite{Browning2019}, as well as analysing travelling wave solutions of these types of multi--species mathematical models~\cite{Colson2021,Elhachem2021,Gallay2001}.

In this work we study a continuum mathematical model of cell invasion that is motivated by the experimental images in Figure \ref{fig:1}(a). The mathematical model explicitly describes the evolution of the cell density, $\hat{u}(\hat{\mathbf{x}},\hat{t})$, and the density of substrate produced by the cells, $\hat{s}(\hat{\mathbf{x}},\hat{t})$, giving rise to a coupled system of nonlinear partial differential equations (PDE).  We first explore numerical solutions of the mathematical model in two spatial dimensions to mimic the same patterns of tissue development that we see in the experimental images in Figure \ref{fig:1}(a).

Within this modelling framework, it is natural for us to ask how the duration of time required for the pore to close is affected by the dynamics of substrate deposition and decay.  We address this question by nondimensionalising the mathematical model, and numerically exploring travelling wave solutions in one dimension.  Not only does travelling wave analysis of the mathematical model have a direct link to the application in question, we note that travelling wave analyses provide detailed mathematical insight into various models of invasion with applications including tissue engineering~\cite{Landman2007}, directed migration~\cite{Krause2020}, disease progression~\cite{Strobl2020} and various applications in ecology~\cite{Hogan2017,Elhachem2021a}.  Our preliminary numerical explorations suggest that, similar to the well--known Porous--Fisher model, the substrate model supports both sharp--fronted and smooth travelling wave solutions. Working in three--dimensional phase space, we show that travelling wave solutions exist for all wave speeds $c \ge c_{\rm{min}}$, where  $c_{\rm{min}} > 0$ is some minimum wave speed, and we provide a geometric argument based on a slow manifold reduction to distinguish between sharp--fronted travelling wave solutions that move with the minimum speed $c_{\rm{min}}$, from smooth travelling wave solutions that move faster than the minimum speed, $c > c_{\rm{min}}$.  The three--dimensional phase space arguments are supported by some analysis of the time--dependent PDE problem where we show how the long--time travelling wave speed relates to the initial decay rate of the cell density.  All phase--space and time--dependent PDE analysis throughout this work are supported by detailed numerical simulations of the full time--dependent PDE model.  For completeness  we also present various perturbation solutions that give accurate mathematical expressions describing the shape of the travelling waves profiles in various limits.

Overall, we show that the substrate invasion model can be viewed as bridge between the relatively simple Porous--Fisher model and more detailed mathematical models of biological invasion.  The substrate model supports various types of travelling wave solutions that are reminiscent of travelling wave solutions of the Porous--Fisher model, but the analysis of these travelling wave solutions is quite different, as we shall now explore.

\section{Results and Discussion}
In this work all dimensional variables and parameters are denoted with a circumflex, and nondimensional quantities are denoted using regular symbols.

\subsection{Biological motivation}\label{sec:biologicalmotivation}
Following Buenzli et al.~\cite{Buenzli2020}, we consider the following minimal model of cell invasion
\begin{align}
\label{eq:GouvdiffUDimensional2D}
&\dfrac{\partial \hat{u}}{\partial \hat{t}}=\hat{D} \div{\left(\dfrac{\hat{s}}{\hat{K}_s} \grad \hat{u}\right)} +\hat{\lambda} \hat{u}\left(1-\dfrac{\hat{u}}{\hat{K}_u}\right),& \hat{\mathbf{x}} \in \Omega,\\
&\dfrac{\partial \hat{s}}{\partial \hat{t}}= \hat{r}_1 \hat{u} - \hat{r}_2 \hat{s},& \hat{\mathbf{x}} \in \Omega, \label{eq:GouvdiffSDimensional2D}
\end{align}
where $\hat{u}(\hat{\mathbf{x}},\hat{t}) \ge 0$ is the density of cells, $\hat{s}(\hat{\mathbf{x}},\hat{t}) \ge 0$ is the substrate concentration, $\hat{D}>0$ is the cell diffusivity and $\hat{\lambda}>0$ is the cell proliferation rate.  This model assumes that cells produce an adhesive and immobile substrate at rate $\hat{r}_1>0$, and that the substrate decays at a rate $\hat{r}_2 > 0$.  We assume that the carrying capacity density of cells is $\hat{K}_u>0$, and that a typical maximum substrate density is $\hat{K}_s>0$.  The key feature of this mathematical model is that the diffusive flux of cells is proportional to the substrate density, $\hat{s}$.  This assumption couples the cell density to the substrate concentration in a way that the diffusive flux vanishes when $\hat{s}=0$.  In this model the evolution of the cell density is affected by the substrate through the cell migration term, without any direct coupling in the cell proliferation term.  This assumption is consistent with recent two--dimensional studies that explored how different surface coatings affect combined cell migration and cell proliferation in wound healing assays~\cite{Jin2020}.  This work showed that different surface coatings have a dramatic impact on cell migration, whereas cell proliferation is less sensitive.

In this modelling framework we make use of the fact that the tissues produced in the experiments in Figure \ref{fig:1}(a) are thin; the horizontal length scale is approximately $300$ $\mu$m whereas the depth of tissue is approximately one cell diameter only, which is around $10-20$ $\mu$m.  In this setting it is appropriate and accurate to use a depth--averaged modelling framework where variations in the vertical direction are implicit, rather than being explicitly described~\cite{Simpson2009}.

We begin by considering Equations (\ref{eq:GouvdiffUDimensional2D})--(\ref{eq:GouvdiffSDimensional2D}) on a two--dimensional square--shaped domain, $\Omega = \{(\hat{x},\hat{y}):0 \le \hat{x} \le \hat{L}, 0 \le \hat{y} \le  \hat{L}\}$ to match the geometry of the experiments in Figure \ref{fig:1}(a).  For simplicity we work with Dirichlet boundary conditions by setting $\hat{u}=\hat{K}_u$ and $\hat{s}= \hat{r}_1 \hat{K}_u / \hat{r}_2$ along all boundaries, with spatially uninform initial conditions $\hat{u} = \hat{s} = 0$, at  $\hat{t}=0$.  A numerical solution of Equation (\ref{eq:GouvdiffUDimensional2D})--(\ref{eq:GouvdiffSDimensional2D}) in Figure \ref{fig:1}(b)--(c) shows the evolution of $\hat{u}$ and $\hat{s}$, respectively.  Full details of the numerical methods used to solve Equations (\ref{eq:GouvdiffUDimensional2D})--(\ref{eq:GouvdiffSDimensional2D}) are given in the Supplementary Material.  The evolution of $\hat{u}$ in Figure \ref{fig:1}(b) shows that the model predicts the sharp--fronted tissue growth that qualitatively matches the spatial and temporal patterns observed in the experiment. The evolution of $\hat{s}$ in Figure \ref{fig:1}(c) shows that the invading cell density profile is associated with an invading  substrate profile.  The coupling between the spatial and temporal distribution of the tissue and the underlying substrate is similar to that observed in the experiments~\cite{Lanaro2021}.  Given this experimental motivation we will now set about analyzing the mathematical model to provide insight into how the substrate dynamics affect the speed of invasion.

\subsection{One-dimensional numerical exploration}\label{sec:1Dexploration}
For the purpose of studying travelling wave solutions of the substrate model we re--write Equations (\ref{eq:GouvdiffUDimensional2D})--(\ref{eq:GouvdiffSDimensional2D}) in the one--dimensional Cartesian coordinate system.  Introducing the following dimensionless quantities: $u=\hat{u} /\hat{K}_u$, $s=\hat{s} /\hat{K}_s$, $x = \hat{x}\sqrt{\hat{\lambda}/\hat{D}}$, $t = \hat{\lambda} \hat{t}$, $r_1 = \hat{r}_1\hat{K}_u/(\hat{\lambda}\hat{K}_s)$ and $r_2 = \hat{r}_2/\hat{\lambda}$, gives the following non--dimensional model
\begin{align}
\label{eq:GouvdiffUNonDimensional}
&\dfrac{\partial u}{\partial t}= \dfrac{\partial}{\partial x}\left(s \dfrac{\partial u}{\partial x}\right) + u(1-u), &0 < x < \infty\\
&\dfrac{\partial s}{\partial t}= r_1 u - r_2 s, &0 < x < \infty, \label{eq:GouvdiffSNonDimensional} \\
&\dfrac{\partial u(0,t)}{\partial x}=0, \qquad \textrm{and} \quad \ u(x,t) \rightarrow 0, \ x \rightarrow \infty. \label{eq:BCNonDimensional}
\end{align}
This dimensionless model involves just two free parameters that relate to the rate of substrate production and the rate of substrate decay, $r_1$ and $r_2$, respectively. Note that Equation (\ref{eq:GouvdiffSNonDimensional}) does not involve any spatial derivatives so there is no need to specify any boundary conditions for $s$.

In this study we will consider two different types of initial conditions: (i) a biologically--realistic initial condition describing the situation where the initial cell population occupies a particular region, and the cell density vanishes outside of this region~\cite{Maini2004,Simpson2013,Sengers2007}; and, (ii) a mathematically insightful, but less biologically--realistic initial condition where the initial cell density decays exponentially as $x \to \infty$.  For the biologically--realistic initial conditions we always consider
\begin{align}
u(x,0) &= 1 - H(\beta), \label{eq:ICUCompactSupport}\\
s(x,0) &= 0,	 \label{eq:ICSCompactSupport}
\end{align}
on $0 < x < \infty$, where $H(x)$ is the usual Heaviside function and $\beta > 0$ is a constant describing the initial length of the domain that is occupied at $t=0$.  For the mathematically interesting initial condition we always consider
\begin{align}
\label{eq:ICUNonCompactSupport}
u(x,0) &=
\begin{cases}
1, & \ x < \beta, \\
\textrm{exp}[-a(x-\beta)], & \ x > \beta,
\end{cases}\\
s(x,0) &= 0,	 \label{eq:ICSNonCompactSupport}
\end{align}
on $0 < x < \infty$, where $a > 0$ is the decay rate. For all results, we set $\beta = 10$.

We focus on long--time numerical solutions of Equations (\ref{eq:GouvdiffUNonDimensional})-- (\ref{eq:GouvdiffSNonDimensional}) in order to explore travelling wave solutions.  Details of the numerical method we use to solve the governing equations are given in the Supplementary Material.   Of course, the travelling wave analysis of this model is relevant on the infinite domain, $0 < x < \infty$, but numerically we must always work with a truncated domain $0 < x < X$, where $X$ is chosen to be sufficiently large that the late--time numerical solutions are unaffected by the choices of $X$.  All algorithms required to re--create the results in this work are available on \href{https://github.com/ProfMJSimpson/Substrate_Mediated_Invasion}{GitHub}.

\begin{figure}[H]
	\centering
	\includegraphics[width=1\linewidth]{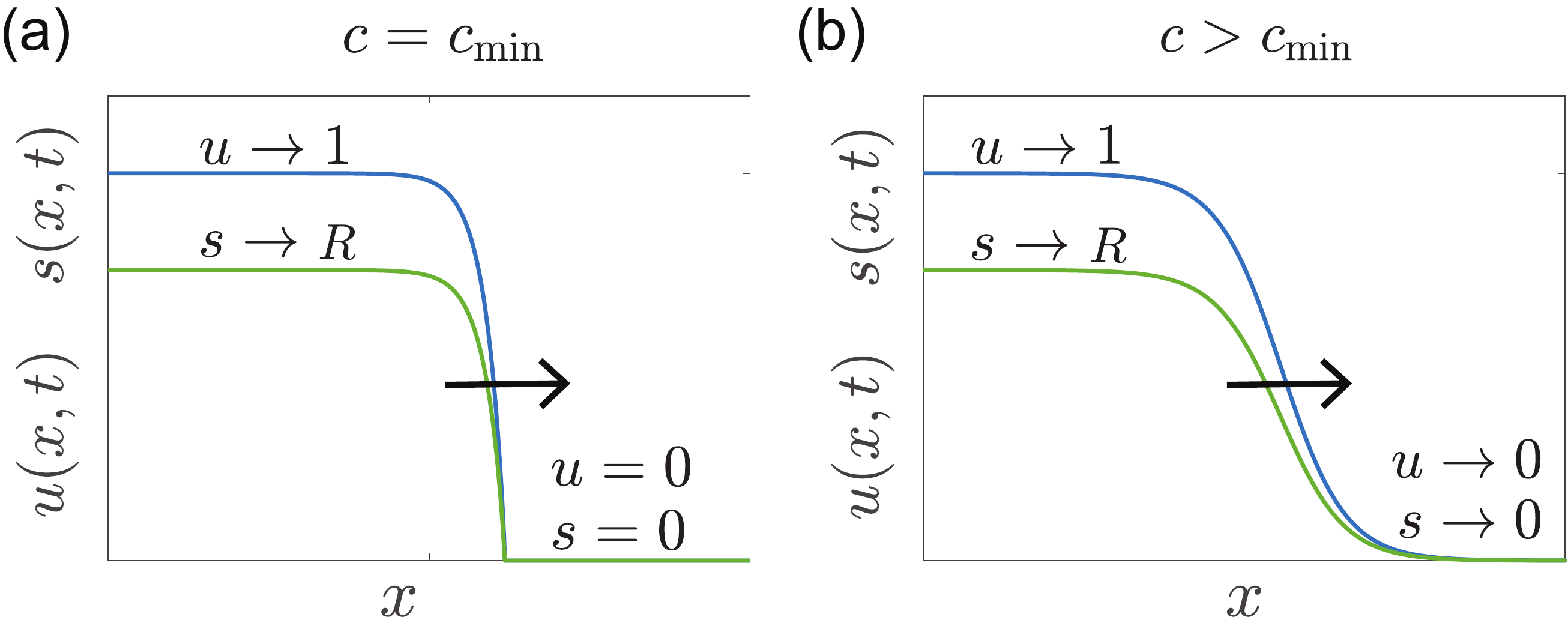}
	\caption{\textbf{Schematic showing sharp and smooth--fronted travelling wave solutions.} (a) Schematic showing a sharp-fronted travelling wave. (b) Schematic showing a smooth-fronted travelling wave.  Arrows show the direction of movement.}
	\label{fig:2}
\end{figure}

Before we present and discuss particular travelling wave solutions, it is convenient to state at the outset that we find the substrate invasion model leads to two types of travelling wave solutions, shown schematically in Figure \ref{fig:2}.  The travelling wave solution in Figure \ref{fig:2}(a) arises from the biologically--relevant initial conditions (\ref{eq:ICUCompactSupport})--(\ref{eq:ICSCompactSupport}), where we see that there is a well--defined sharp front with $u=s=0$ ahead of the front, and $u \to 1^-$ and $s \to R^-$ well--behind the travelling wave front as $x \to 0^-$.  In this case, as we will show, the travelling wave solution corresponds to the minimum wave speed, $c = c_{\rm{min}}$, that depends on the value of $r_1$ and $r_2$. In contrast, the travelling wave solution in Figure \ref{fig:2}(b) arises from the mathematically interesting initial conditions (\ref{eq:ICUNonCompactSupport})--(\ref{eq:ICSNonCompactSupport}).  In this second type of travelling wave we have the same behaviour well--behind the wave front as in Figure \ref{fig:2}(a), since $u \to 1^-$ and $s \to R^-$ as $x \to 0^-$.  However, in this case we have a smooth travelling wave with $u \to 0^+$ and $s \to 0^+$ as $x \to \infty$.  Further, as we will show, these smooth--fronted travelling wave solutions move with a faster travelling wave speed, $c > c_{\rm{min}}$.

The fact that the substrate model gives rise to both smooth and sharp--fronted travelling wave solutions is very interesting and worthy of exploration.  Throughout this work we will explore parallels between the substrate model and the Porous--Fisher model, and an obvious point of similarity is that both these models support smooth and sharp--fronted travelling wave solutions~\cite{Murray2002,Sanchez1994,Sherratt1996}.  As we will explore in this work, however, the differences between the smooth and sharp--fronted travelling waves in the substrate model are more subtle than the Porous--Fisher model, and we must use different methods of analysis to understand these differences.

\begin{landscape}
	\begin{figure}[h!]
		\centering
		\includegraphics[width=1\linewidth]{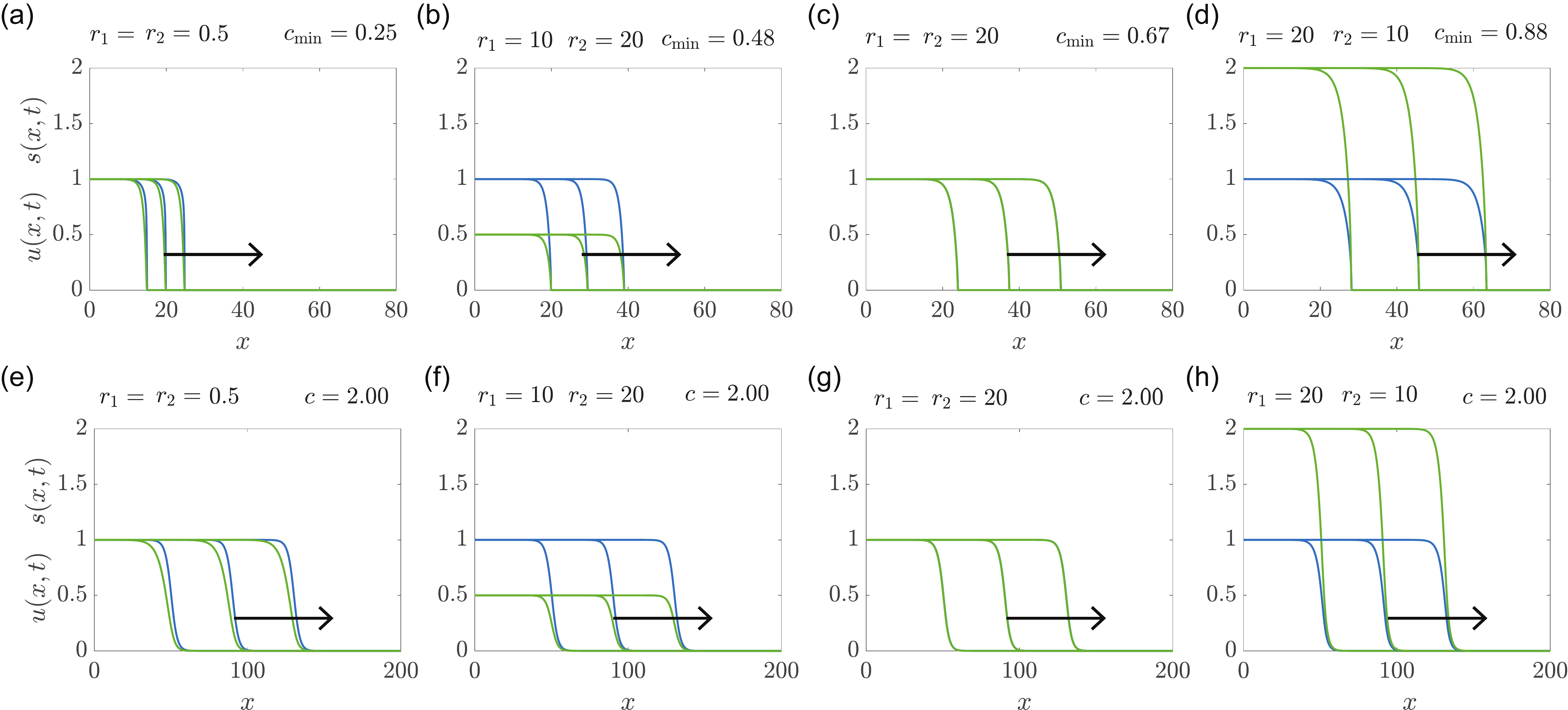}
		\caption{\textbf{Time-dependant PDE solutions showing smooth and sharp--fronted travelling wave solutions.}  Sharp-fronted travelling wave solutions in (a)--(d) are obtained by solving Equations (\ref{eq:GouvdiffUNonDimensional})--(\ref{eq:BCNonDimensional}) with (\ref{eq:ICUCompactSupport})--(\ref{eq:ICSCompactSupport}).  Smooth--fronted travelling wave solutions in (e)--(h) are obtained by solving Equations (\ref{eq:GouvdiffUNonDimensional})--(\ref{eq:BCNonDimensional}) with (\ref{eq:ICUNonCompactSupport})--(\ref{eq:ICSNonCompactSupport}) and $a=1/2$.  Values of $r_1$ and $r_2$ are indicated on each subfigure, and the long-time estimate of the travelling wave speed $c$ is also given to two decimal places.  Each subfigure shows profiles for $u(x,t)$ (blue) and $s(x,t)$ (green) at $t=20, 40$ and $60$, with the arrow showing the direction of increasing $t$.  All numerical solutions correspond to  $\Delta x = 1\times10^{-2}$, $\Delta t = 1\times10^{-3}$ and $\epsilon=1\times10^{-10}$.}
		\label{fig:3}
	\end{figure}
\end{landscape}

In addition to the schematic solutions in Figure \ref{fig:2}, we present a range of time--dependent PDE solutions in Figure \ref{fig:3} where we explore the role of varying the substrate dynamics by choosing different values of $r_1$ and $r_2$.
Results in Figure \ref{fig:3}(a)--(c) for the sharp--fronted travelling wave solutions show that the long--time minimum travelling wave speed, $c_{\rm{min}}$, depends on $r_1$ and $r_2$.  In particular, comparing the results in (a)--(d) show that $c_{\rm{min}}$ appears to increase with $r_1$.  In contrast, the smooth--fronted travelling wave solutions in Figure \ref{fig:3}(e)--(h) lead to travelling wave solutions where the wave speed $c > c_{\rm{min}}$ appears to be independent of $r_1$ and $r_2$.  These numerical solutions show that the value of $s$ well--behind the travelling wave front depends on the choice of $r_1$ and $r_2$, and motivates us to define
\begin{equation}
R = \dfrac{r_1}{r_2},
\end{equation}
so that we have $s \to R^-$ as $x \to 0^-$, which is consistent with the schematics in Figure \ref{fig:2}.

\begin{figure}[H]
	\centering
	\includegraphics[width=1\linewidth]{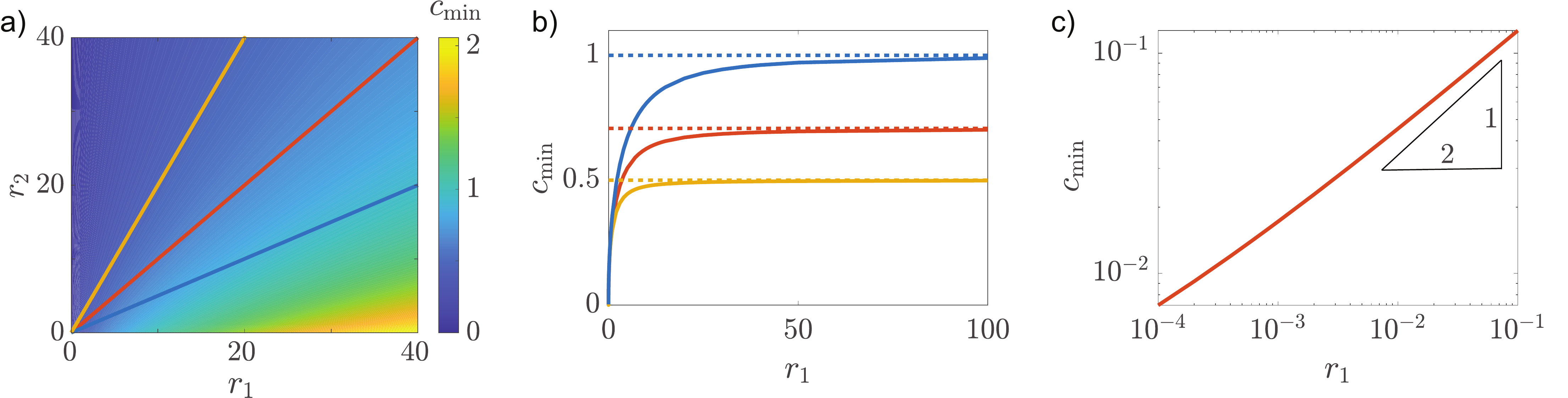}
	\caption{\textbf{Numerical exploration of the relationship between $c_{\rm{min}}$, $r_1$ and $r_2$.} (a) heat map of $c_{\rm{min}}$ as a function of $r_1$ and $r_2$ obtained by solving (\ref{eq:GouvdiffUNonDimensional})--(\ref{eq:BCNonDimensional}) with (\ref{eq:ICUCompactSupport})--(\ref{eq:ICSCompactSupport}).  The three straight lines superimposed on (a) correspond to $R=0.5$ (yellow), $R=1$ (red) and $R=2$ (blue), and the relationship between $c_{\rm{min}}$ and $r_1$ for these fixed values of $R$ is given in (b), showing that $c_{\rm{min}} \to \sqrt{R/2}^{\,-}$ as $r_1 \to \infty$.  (c) shows $c_{\rm{min}}$ as a function of $r_1$ for $R=0.5,1$ and $2$, suggesting that $c_{\rm{min}} \sim A \sqrt{r_1}$ as $r_1 \to 0$, for some constant $A>0$.  All numerical solutions correspond to  $\Delta x = 1\times10^{-2}$, $\Delta t = 1\times10^{-2}$ and $\epsilon=1\times10^{-10}$.}
	\label{fig:4}
\end{figure}

Now we have established that the long--time travelling wave speed for the sharp--fronted travelling wave solutions depends upon $r_1$ and $r_2$, we generate a suite of sharp--fronted travelling wave solutions numerically, and estimate $c_{\rm{min}}$ as a function of $r_1$ and $r_2$, as reported in Figure \ref{fig:4}(a).   This heat map suggests that holding $r_2$ constant and increasing $r_1$ leads to an increase in $c_{\rm{min}}$.  In contrast, holding $r_1$ constant and increasing $r_2$ reduces $c_{\rm{min}}$.  To further explore this relationship we superimpose three straight lines on the heat map in Figure \ref{fig:4}(a).  These straight lines correspond to $R=0.5$ (yellow), $R=1$ (red) and $R=2$ (blue).  Plotting $c_{\rm{min}}$ as a function of $r_1$ for these three fixed values of $R$ in Figure \ref{fig:4}(b) suggest that $c_{\rm{min}} \to \sqrt{R/2}^{\,-}$ for fixed $R$, as $r_1 \to \infty$.  As we will explain later in Section {\ref{sec:approximationr1andr2big}}, this numerical observation is related to the fact that the substrate model simplifies to the Porous--Fisher model when $r_1$ and $r_2$ are sufficiently large~\cite{Buenzli2020}.

Results in Figure \ref{fig:4}(b) explore the fast substrate production limit, $r_1 \to \infty$ for fixed $R$, whereas results in Figure \ref{fig:4}(c) explore the small substrate production limit, $r_1 \to 0$. In this case we plot $c_{\rm{min}}$ as a function for $r_1$, for $R=0.5, 1$ and $2$, and we see that the results for different values of $R$ are identical, suggesting that $c_{\rm{min}}$ is independent of $r_2$ as $r_1 \to 0$.  Furthermore, the straight line relationship on the log--log plot in Figure \ref{fig:4}(c) suggests that we have $c_{\rm{min}} \sim A \sqrt{r_1}$ as $r_1 \to 0$ for some constant $A>0$.

In summary, results in Figure \ref{fig:4} summarise the numerically--determined relationship between $c_{\rm{min}}$, $r_1$, and $r_2$ for sharp--fronted travelling wave solutions of the substrate model.  These numerical results are of interest because some results are consistent with well--known results for the Porous--Fisher model as we further explore in Section \ref{sec:approximationr1andr2big}.  In contrast, we also observe different behaviour that is inconsistent with the Porous--Fisher model.  For example, the non-dimensional Porous-Fisher model has a positive minimum wavespeed $c_{\rm{min}} = 1/\sqrt{2} \approxeq 0.71$, whereas the substrate--mediated invasion model supports sharp--fronted travelling wave solutions with vanishingly small minimum wave speed, $c_{\rm{min}} \to 0$ as $r_1 \to 0$.  Table (\ref{tab:ComparisonPorousFSubstrateModel}) summarises the differences and similarities between travelling wave solutions of the Porous--Fisher model and the substrate model.  While some of these results have only been numerically explored so far, in later sections we will provide more thorough evidence to support these numerically--based observations.

\begin{table}[h!]
	\caption{Key features of travelling wave solutions of the substrate--mediated invasion model with travelling wave solutions of the Porous--Fisher model.}	
		\centering
		\begin{tabular}{|c|c|c|c|}
			\hline
			\multicolumn{2}{|c|}{Porous-Fisher} & \multicolumn{2}{|c|}{Substrate--mediated model}\\
			\hline
			{Smooth front} & {Sharp front} & {Smooth front}  & {Sharp front} \\
			\hline
			$\begin{aligned}
			c = \begin{cases} \dfrac{1}{a} & a < \sqrt{2}\\ \dfrac{1}{\sqrt 2} & a \ge \sqrt{2}\end{cases}
			\end{aligned}$ 	& $\begin{aligned}
			c_{\textrm{min}} = {\dfrac{1}{\sqrt 2}}
			\end{aligned}$ & $\begin{aligned}
			c &= \dfrac{1}{a} \\
			\lim_{\substack{r_1 \to \infty \\ r_2 \to \infty}} c &= {\sqrt{\dfrac{R}{2}}}^{\ -}
			\end{aligned}$ &   $\begin{aligned}
			\lim_{r_1 \to 0^+}c_{\textrm{min}} &= 0^+\\
			\lim_{\substack{r_1 \to \infty \\ r_2 \to \infty}}c_{\textrm{min}} &= {\sqrt{\dfrac{R}{2}}}^{\ -}
			\end{aligned}$ \\
			\hline
		\end{tabular}
		\label{tab:ComparisonPorousFSubstrateModel}
\end{table}
Given the numerical evidence developed in this section, we will now use phase space techniques to understand the differences between the sharp--fronted and smooth--fronted travelling wave solutions of the substrate model.

\subsection{Phase space analysis for smooth travelling wave solutions}\label{sec:Phaseplaneinz}
In the usual way, we seek to study travelling wave solutions of Equations (\ref{eq:GouvdiffUNonDimensional})--(\ref{eq:GouvdiffSNonDimensional}) by writing $u(x, t)= U(z)$ and $S(x, t)=S(z)$, where $z$ is the travelling wave variable, $z=x-ct$~\cite{Murray2002} to give
\begin{align}
\dfrac{\mathrm{d}}{\mathrm{d}z}\left(S \dfrac{\mathrm{d}U}{\mathrm{d}z}\right) + c\dfrac{\mathrm{d}U}{\mathrm{d}z} + U(1-U) &= 0,&-\infty < z < \infty, \label{eq:ODEUz}\\
c\dfrac{\mathrm{d} S}{\mathrm{d} z} + r_1 U- r_2 S&= 0,&-\infty < z < \infty. \label{eq:ODESz}
\end{align}
Boundary conditions for the smooth travelling wave solutions are $U(z) \rightarrow 1$ and $S(z) \rightarrow R$ as $z\rightarrow -\infty$, and $U(z) \rightarrow 0$ and $S(z) \rightarrow 0$ as $z\rightarrow \infty$.  Given such a smooth--fronted travelling wave solution for $U(z)$, we can solve Equation (\ref{eq:ODESz}) to give
\begin{equation}
S(z)=\frac{r_1}{c}\textrm{exp}\left[\dfrac{r_2 z}{c}\right] \int_{z}^{\infty} \textrm{exp}\left[\dfrac{-r_2 y}{c}\right] \, U(y) \, \mathrm{d} y. \label{eq:SzasfunctionIntU}
\end{equation}
We will make use of this result later.

Following the usual approach to studying smooth travelling wave solutions, we re--write Equations (\ref{eq:ODEUz})--(\ref{eq:ODESz}) as a first order system
\begin{align}
\label{eq:ODEdUdz}
\dfrac{\mathrm{d} U}{\mathrm{d} z} &= W,\\
\dfrac{\mathrm{d} S}{\mathrm{d} z} &= -\left(\dfrac{r_1 U - r_2 S}{c}\right), \label{eq:ODEdSdz}\\
\dfrac{\mathrm{d} W}{\mathrm{d} z} &=  W\left(\dfrac{r_1 U - r_2 S-c^2}{cS}\right) - \dfrac{ U(1-U)}{S}. \label{eq:ODEdWdz}
\end{align}
There are two equilibrium points of the phase space: (i) $(\bar{U},\bar{S},\bar{W}) = (1,R,0)$ as $z\rightarrow -\infty$, which corresponds to the invaded boundary; and, (ii) $(\bar{U},\bar{S},\bar{W}) = (0,0,0)$ as $z\rightarrow \infty$, which corresponds to the uninvaded boundary.

To explore the possibility of a heteroclinic orbit connecting the two equilibrium points in the three--dimensional phase space, the Jacobian of this system is
\begin{align}
\begin{bmatrix}
0  & 0 & 1\\[5pt]
-\dfrac{r_1}{c}&\dfrac{r_2}{c}& 0\\[5pt]
\displaystyle \dfrac{r_1 \bar{W} - c(1 - 2\bar{U})}{c \bar{S}} & \dfrac{(-r_1 \bar{U} + c^2)\bar{W}+c\bar{U}(1-\bar{U})}{c\bar{S}^2}& \dfrac{ -r_2 \bar{S} +r_1 \bar{U} - c^2}{c\bar{S}}\\[5pt]
\end{bmatrix}.\label{eq:JacobianZ}
\end{align}
We see immediately that we cannot follow the usual practice of evaluating the Jacobian at the uninvaded equilibrium point since it is not defined at $(\bar{U},\bar{S},\bar{W}) = (0,0,0)$ and so linearisation is not useful here.  In contrast, the Jacobian at the invaded equilibrium point  $(\bar{U},\bar{S},\bar{W}) = (1,R,0)$ is
\begin{align}
\begin{bmatrix}
0  & 0 & 1\\
-\dfrac{r_1}{c}&\dfrac{r_2}{c}& 0\\
\dfrac{r_2}{r_1}& 0& -\dfrac{c r_2}{r_1}\\
\end{bmatrix}.\label{eq:JacobianZforsaddle}
\end{align}
The eigenvalues of this Jacobian are $\lambda_1 = r_2/c$ and $\lambda_{2,3} = (-c \pm \sqrt{c^2 + 4 R})/(2R)$.  Since these eigenvalues are all real valued, with $\lambda_{1,2} > 0$ and $\lambda_3 < 0$, the invaded equilibrium point is a three--dimensional saddle point.

As just mentioned, linearisation about the uninvaded equilibrium point is not possible, and so we revisit the dynamical system (\ref{eq:ODEdUdz})--(\ref{eq:ODEdWdz}) as $z \to \infty$ in more detail in Section \ref{sec:slowmanifold} below.  For now, we suppose that a smooth travelling wave $U(z)$ decays exponentially, say
\begin{equation}
	U(z) \sim C\textrm{exp}\left( -b z \right) \quad z \to \infty,  \label{eq:Ulargez}
\end{equation}
where $b > 0$.  Under this assumption it follows from (\ref{eq:SzasfunctionIntU}) that
\begin{align}\
S(z)& \sim \dfrac{r_1}{b c + r_2}U(z),\label{eq:Slargez}\\
W(z)&\sim -b U(z),\label{eq:Wlargez}
\end{align}
suggesting that $S(z)$ and $W(z)$ both decay to zero exponentially, at the same rate as $U(z)$, as $z \to \infty$.  Further, to leading order as $z \to \infty$, (\ref{eq:ODEdWdz}) gives
\begin{align}
\dfrac{\mathrm{d} W}{\mathrm{d} z} &\sim (bc-1)\left(\dfrac{b c + r_2}{r_1}\right) \quad \textrm{as} \quad z \to \infty.
\end{align}
At first glance this results appears inconsistent with our arguments so far, since for smooth travelling wave solutions we expect $\mathrm{d} W / \mathrm{d} z \to 0$ as $z \to \infty$, but here we have $\mathrm{d} W / \mathrm{d} z$ approaching a constant.  However, by choosing  $c=1/b$ we avoid this inconsistency.  This choice implies that the speed of the smooth--fronted travelling wave is related to the far--field decay rate of $U(z)$.  We have tested this hypothesis numerically and found and excellent match between (\ref{eq:Ulargez})--(\ref{eq:Wlargez}) and the shape of the smooth--fronted travelling waves for different choices of $r_1$, $r_2$ and $c$, with one example discussed in the Supplementary Material.  In addition, we provide further evidence for this far--field behaviour in Section \ref{sec:slowmanifold}.

\subsection{Dispersion relationship}\label{sec:dispersionrelationship}
We now explore how the decay rate of the initial condition, $a$ in Equation (\ref{eq:ICUNonCompactSupport}), affects the long--time travelling wave speed for smooth--fronted travelling wave solutions.  To be consistent with  our observations in Section \ref{sec:Phaseplaneinz}, we assume that smooth--fronted travelling wave solutions for $U(z)$ and $S(z)$ decaying at the same rate, and we seek solutions of the form $\tilde{u}(x,t) \sim C \ \textrm{exp} \ [a(x-ct)]$ and $\tilde{s}(x,t) \sim D \ \textrm{exp} \ [a(x-ct)]$ as $x \to \infty$.  Substituting these solutions into Equation (\ref{eq:GouvdiffUNonDimensional}), and focusing on the leading edge of these solutions where $\tilde{u}(x,t) \ll 1$, we obtain
\begin{equation}
c =  \dfrac{1}{a}, \label{eq:DispersionRelationship}
\end{equation}
which relates the long--time speed of the travelling wave solution to the decay rate of the initial condition, $u(x,0)$.

\begin{landscape}
\begin{figure}[H]
	\centering
	\includegraphics[width=1\linewidth]{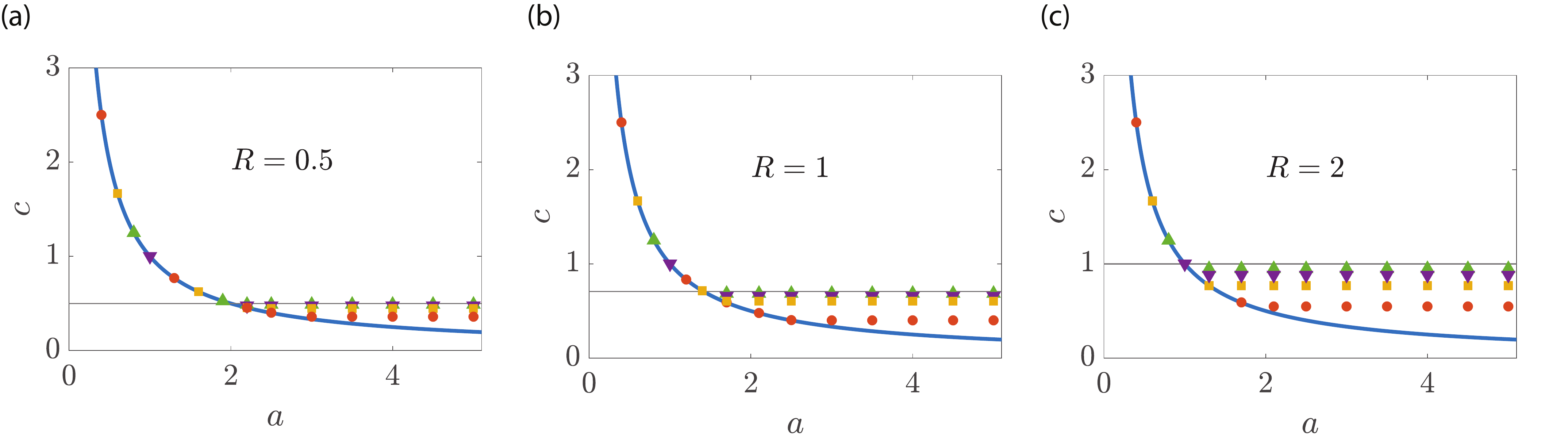}
	\caption{\textbf{Dispersion relationship.} (a)--(c) shows $c$ as a function of the initial decay rate, $a$, for $R=0.5, 1$ and $2$, respectively. Numerical travelling wave speeds are estimated from long--time numerical solutions of Equations (\ref{eq:GouvdiffUNonDimensional})--(\ref{eq:BCNonDimensional}) with the initial condition given by Equations  (\ref{eq:ICUNonCompactSupport})--(\ref{eq:ICSNonCompactSupport}) with various values of $a$. The dispersion relationship, Equation (\ref{eq:DispersionRelationship}), is plotted (solid blue) and results for $r_1 = 1, 5, 10$ and $20$ are shown in orange discs, yellow squares, purple triangles and green triangles, respectively.   Each plot shows a horizontal line at $\sqrt{R/2}$, which is an upper bound for the wavespeed for large $a$.  All numerical PDE solutions correspond to  $\Delta x = 1\times10^{-2}$, $\Delta t = 1\times10^{-3}$ and $\epsilon=1\times10^{-10}$.}
	\label{fig:5}
\end{figure}
\end{landscape}

Results in Figure \ref{fig:5} explore the validity of Equation (\ref{eq:DispersionRelationship}) by taking time--dependent PDE solutions with initial conditions (\ref{eq:ICUNonCompactSupport})--(\ref{eq:ICSNonCompactSupport}) and varying the decay rate of $u(x,0)$ for various values of $r_1$ and $r_2$.  In particular, we generate travelling wave solutions for $r_1 = 1, 5, 10$ and $20$,  for fixed $R=0.5, 1$ and $2$.  Results in Figure \ref{fig:5}(a)--(c) corresponding to $R=0.5, 1$ and 2, respectively, show that for sufficiently small $a$, we see that the long--time travelling wave speed matches Equation (\ref{eq:DispersionRelationship}) regardless of $r_1$ and $r_2$.  These results are consistent with the initial explorations in Figure \ref{fig:3}(e)--(h) where we saw that the wave speed of certain smooth--fronted travelling wave solutions was independent of $r_1$ and $r_2$.  As $a$ increases, however, we see that $c$ behaves differently.  For large $a > a_{\textrm{crit}}$ we see that $c$ approaches a constant value $c_{\textrm{min}}$ that is  independent of $a$.  Our numerical evidence suggests that this limiting constant value depends on $r_1$ and $r_2$.  For completeness, on each subfigure we plot a horizontal line at $c = \sqrt{R/2}$, and we note that this value appears to be an upper--bound for $c$ as $a$ becomes large.

\begin{figure}[H]
	\centering
	\includegraphics[width=0.9\linewidth]{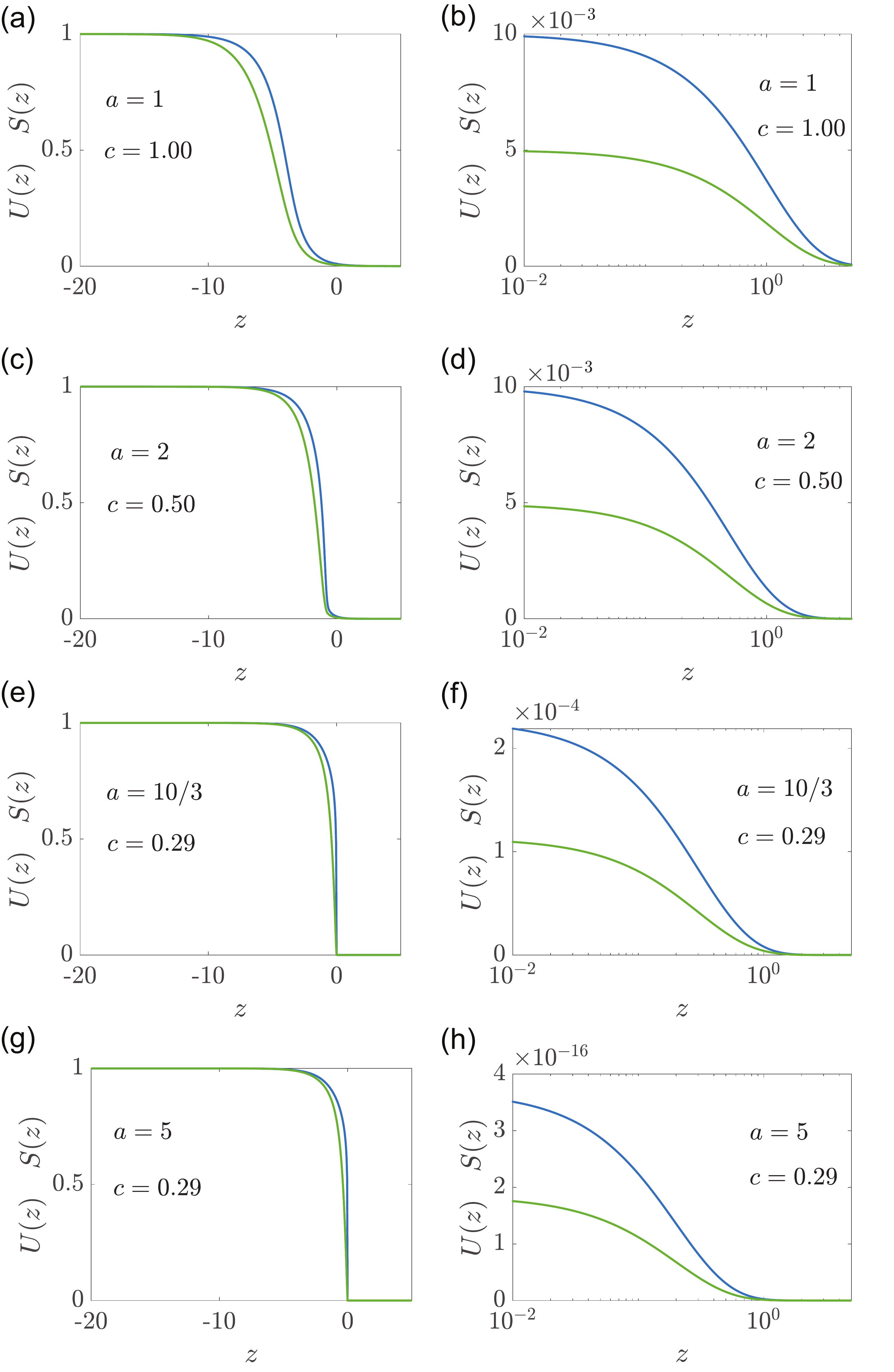}
	\caption{\textbf{Smooth-fronted travelling wave solutions.}  Travelling wave solutions $U(z)$ and $S(z)$ are obtained by considering long--time numerical solutions of Equations (\ref{eq:GouvdiffUNonDimensional})--(\ref{eq:BCNonDimensional}) with initial conditions given by Equations  (\ref{eq:ICUNonCompactSupport})--(\ref{eq:ICSNonCompactSupport}) with variable decay rate, $a$.  All results correspond to $r_1=r_2=1$, and results in (a)--(b), (c)--(d), (e)--(f) and (g)--(h) correspond to $a=1, 2, 10/3$ and $5$, respectively, as indicated.  Results in the left--most column show the various travelling wave solutions plotted on the usual scale with $0 \le U(z), S(z) \le 1$.  Results in the right--most column show a magnification of the leading edge of the travelling waves.}
	\label{fig:6}
\end{figure}

The transition from $c = 1/a$ for $a < a_{\textrm{crit}}$ to constant $c$ for $a > a_{\textrm{crit}}$ in Figure \ref{fig:5} is further explored in Figure \ref{fig:6} for $r_1=r_2=1$.  The long--time travelling wave solution in Figure \ref{fig:6}(a)--(b) evolves from an initial condition with decay rate $a=1$. This solution evolves into a smooth travelling wave with $c=1.00$, which is consistent with the dispersion relationship, Equation (\ref{eq:DispersionRelationship}).  Although it is clear that the travelling wave solution in Figure \ref{fig:6}(a) is smooth at this scale, we also plot a magnification of the leading edge of that travelling wave in Figure \ref{fig:6}(b).   We now explore a series of travelling wave solutions as $a$ increases to visualise the transition reported in Figure \ref{fig:5}. The long--time travelling wave solution in Figure \ref{fig:6}(c)--(d) evolves from an initial condition with a faster decay rate, $a=2$, leading to a smooth--fronted travelling wave with $c=0.50$. Again, this result is consistent with the dispersion relationship, and the magnification of the density profiles near the leading edge in Figure \ref{fig:6}(d) confirms that the travelling wave solution is smooth.  The travelling wave solution in Figure \ref{fig:6}(e) for $a=10/3$ leads to a travelling wave solution with $c=0.29$.  This estimate from the long--time numerical solution of the PDE is close to the travelling wave speed predicted by the dispersion relationship.  At the scale shown in Figure \ref{fig:6}(e) it might seem, at first glance, that the travelling wave is sharp, but the magnification in Figure \ref{fig:6}(f) confirms that this travelling wave is indeed smooth--fronted.   Finally, the travelling wave solution in Figure \ref{fig:6}(g) for $a=5$ evolves to a travelling wave solution with $c=0.29$, which is much larger than the speed predicted by the dispersion relationship that would give $c = 1/5 = 0.2$.  Again, while the travelling wave solution in Figure \ref{fig:6}(g) appears to be sharp at this scale, the magnification of the solution in Figure \ref{fig:6}(h) confirms that this solution is indeed smooth--fronted.

In summary, the dispersion relationship suggests that long--time speed of smooth--fronted travelling wave solutions is given by $c = 1/a$, where $a$ is far-field the decay rate of $u(x,0)$.  Our numerical explorations in Figures \ref{fig:5}--\ref{fig:6} confirms that this result holds for sufficiently small decay rates, $a < a_{\textrm{crit}}$.  As the decay rate increases, $a >  a_{\textrm{crit}}$, we observe an interesting transition for  smooth--fronted travelling waves where $c$ becomes independent of $a$, and greater than the speed predicted by the dispersion relationship.  While these travelling wave solutions remain smooth--fronted as $a$ increases, it becomes increasingly difficult to draw a visual distinction between these smooth--fronted travelling wave solutions and sharp--fronted travelling wave solutions that evolve from initial conditions with compact support, such as those travelling waves in Figure \ref{fig:3}(a)--(d).   We now seek to provide a geometric interpretation of the differences between these two classes of travelling wave solutions by returning to the phase space.

\subsection{Desingularised phase space and slow manifold reduction}\label{sec:slowmanifold}
We now return to the phase space for travelling wave solutions and introduce a change of variables
\begin{equation}
\zeta(z) = \int_{0}^{z} \dfrac{\mathrm{d}y}{S(y)}, \label{eq:Transformation}
\end{equation}
which removes the singularity in Equation (\ref{eq:ODEdWdz}) when $S(z)=0$.  A similar transformation to desingularise the phase plane is often used in the analysis of sharp--fronted travelling wave solutions of the Porous--Fisher model~\cite{Murray2002}.  The desingularised system is given by
\begin{align}
\label{eq:ODEdUdzeta}
\dfrac{\mathrm{d} U}{\mathrm{d} \zeta} &= SW,\\
\dfrac{\mathrm{d} S}{\mathrm{d} \zeta} &= -S\left(\dfrac{r_1 U - r_2 S}{c}\right), \label{eq:ODEdSdzeta}\\
\dfrac{\mathrm{d} W}{\mathrm{d} \zeta} &= W\left(\dfrac{r_1 U - r_2 S-c^2}{c}\right)-  U(1-U). \label{eq:ODEdWdzeta}
\end{align}
There are two equilibrium points of the desingularised phase space: (i) $(\bar{U},\bar{S},\bar{W}) = (1,R,0)$  as $\zeta \to  -\infty$, corresponding to the invaded boundary; and, (ii) $(\bar{U},\bar{S},\bar{W}) = (0,0,0)$ as $\zeta \to \infty$, corresponding to the uninvaded boundary.  It is important to point out that the phase space analysis in Section \ref{sec:Phaseplaneinz} was relevant only for smooth--fronted travelling wave solutions, whereas the desingularised phase space is appropriate for both the sharp--fronted and smooth--fronted travelling wave solutions.  The Jacobian of this system is
\begin{align}
\begin{bmatrix}
0  & \bar{W} & \bar{S}\\[5pt]
-\dfrac{r_1 \bar{S}}{c}&\dfrac{-r_1 \bar{U} + 2 r_2 \bar{S}}{c}& 0\\[5pt]
\dfrac{r_1 \bar{W} - c(1 - 2\bar{U})}{c} & -\dfrac{r_2\bar{W}}{c}& \dfrac{ -r_2 \bar{S} +r_1 \bar{U} - c^2}{c}\\[5pt]
\end{bmatrix}.\label{eq:JacobianInZeta}
\end{align}
We can now consider both equilibrium points $(\bar{U}, \bar{S}, \bar{W}) = (1,R,0)$ and $(\bar{U}, \bar{S}, \bar{W}) = (1,0,0)$.

The Jacobian at the invaded equilibrium point, $(\bar{U}, \bar{S}, \bar{W}) = (1,R,0)$, is
\begin{align}
\begin{bmatrix}
0  & 0 & \dfrac{r_1}{r_2}\\
-\dfrac{r_1^2}{r_2}&\dfrac{r_1}{c}& 0\\
1& 0& -c\\
\end{bmatrix}.\label{eq:JacobianZetaSaddle}
\end{align}
The eigenvalues of this Jacobian are $\lambda_1 = r_1/c$ and $\lambda_{2,3} = (-c \pm \sqrt{c^2 + 4 R})/2$.  Since $\lambda_{1,2} > 0$ and  $\lambda_3 < 0$, the uninvaded equilibrium point is a three--dimensional saddle.   These expressions are identical to the corresponding expressions in Section (\ref{sec:Phaseplaneinz}), which is not surprising since $\zeta = z$ near the invaded equilibrium point, $z \to -\infty$.

\begin{figure}[H]
	\centering
	\includegraphics[width=1\linewidth]{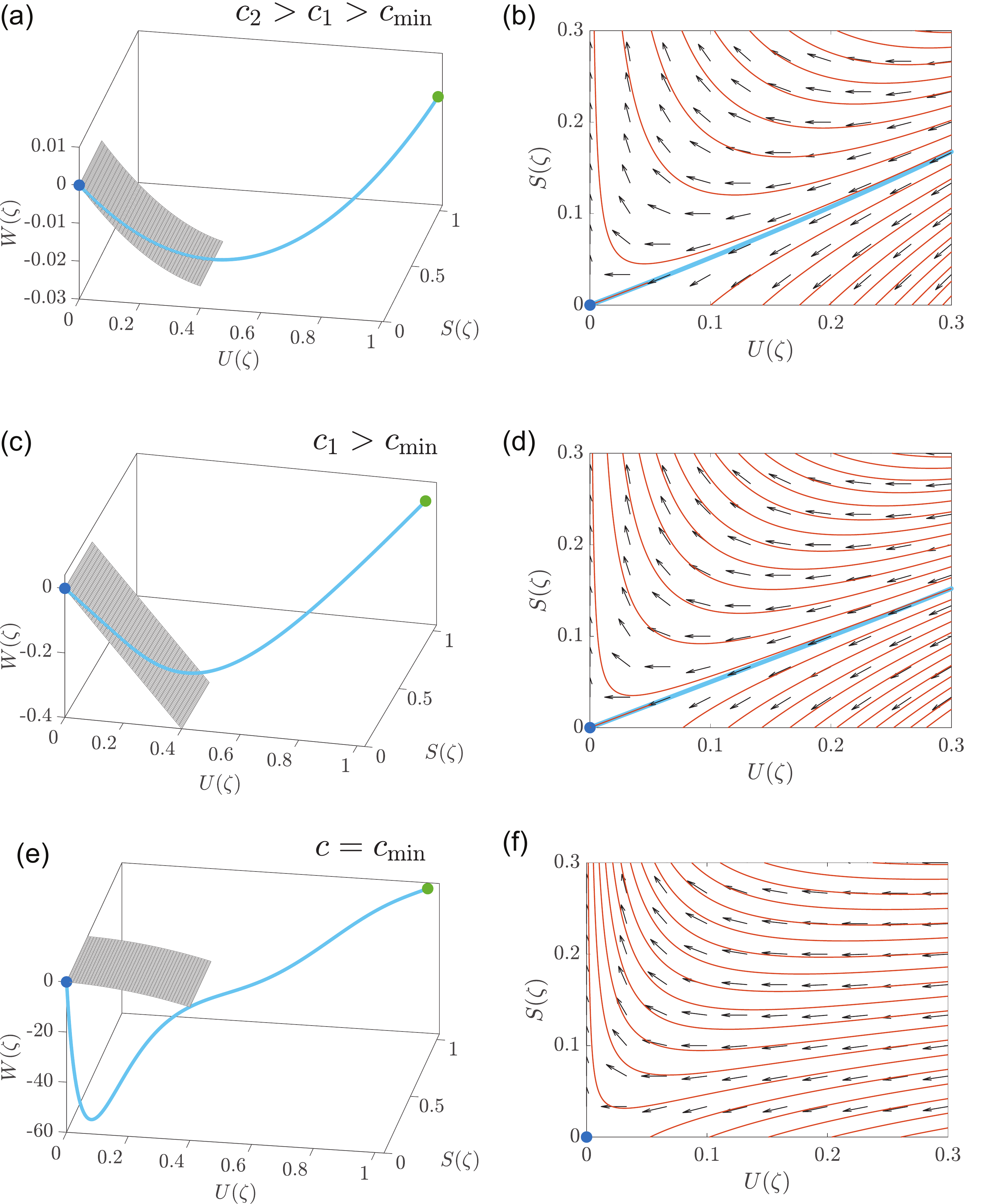}
	\caption{\textbf{Desingularised phase space and slow manifold reduction}. All results correspond to $r_1=r_2=1$. Results in: (a)--(b) correspond to a smooth--fronted travelling wave with $c_2=10$; (c)--(d) correspond to a smooth--fronted travelling wave with $c_1=1$; and, (e)--(f) correspond to a sharp--fronted travelling wave with $c_{\rm{min}} = 0.29$.  Results in the left--most column show the three--dimensional desingularised phase space with the invaded equilibrium point (green dot), the uninvaded equilibrium point (blue dot) and the slow manifold (grey surface).  Results in the right--most column show the vector field on the slow manifold, superimposed with several solution trajectories, including the heteroclinic orbit (blue) and several unphysical trajectories (red).  The heteroclinic orbit is obtained by solving Equations (\ref{eq:GouvdiffUNonDimensional})--(\ref{eq:BCNonDimensional}) numerically with appropriate initial conditions.  For (a)--(b) and (c)--(d) the initial conditions are given by Equations (\ref{eq:ICUNonCompactSupport})--(\ref{eq:ICSNonCompactSupport}) with $a=1/10$ and $a=1$, respectively.  For (e)--(f) the initial conditions are given by Equations (\ref{eq:ICUCompactSupport})--(\ref{eq:ICSCompactSupport}). All numerical PDE solutions correspond to  $\Delta x = 1\times10^{-4}$, $\Delta t = 1\times10^{-3}$ and $\epsilon=1\times10^{-4}$.}	
	\label{fig:7}
\end{figure}

The Jacobian at the uninvaded equilibrium point, $(\bar{U}, \bar{S}, \bar{W}) = (0,0,0)$, is
\begin{align}
\begin{bmatrix}
0  & 0 & 0\\
0&0& 0\\
-1& 0& -c\\
\end{bmatrix}.\label{eq:JacobianZetaHyperbolic}
\end{align}
The eigenvalues are $\lambda_1 = -c$ and $\lambda_2 = \lambda_3 = 0$, which means that $(\bar{U}, \bar{S}, \bar{W}) = (0,0,0)$ is a non-hyperbolic equilibrium point suggesting that the dynamics near this point take place on a slow manifold~\cite{Wiggins2003}.  To explore these local dynamics near $(\bar{U}, \bar{S}, \bar{W}) = (0,0,0)$ we apply the centre manifold theory to identify the slow manifold.  To proceed we rotate the coordinate system using a transformation defined by the eigenvectors $[-c,0,1]^\top$, $[0,1,0]^\top$ and $[0,0,1]^\top$ that are associated with $\lambda_1$, $\lambda_2$ and $\lambda_3$, respectively.  The relationship between the original unrotated coordinate system $(U,S,W)$ and the rotated coordinate system $(\mathscr{U},\mathscr{S},\mathscr{W})$ is given by the transformation~\cite{Maclaren2020},
\begin{align}
	\begin{bmatrix}
		U\\
		S\\
		W\\
	\end{bmatrix}
=
	\begin{bmatrix}
		-c  & 0 & 0\\
		0&1& 0\\
		1& 0& 1\\
	\end{bmatrix}
	\begin{bmatrix}
	\mathscr{U}\\
	\mathscr{S}\\
	\mathscr{W}\\
	\end{bmatrix},\label{eq:eigenspaceTransformation}
\end{align}
and the associated inverse transformation
\begin{align}
	\begin{bmatrix}
		\mathscr{U}\\
		\mathscr{S}\\
		\mathscr{W}\\
	\end{bmatrix}
	= \dfrac{1}{c}
	\begin{bmatrix}
		-1  & 0 & 0\\
		0&c& 0\\
		1& 0& c\\
	\end{bmatrix}
	\begin{bmatrix}
		U\\
		S\\
		W\\
	\end{bmatrix}.\label{eq:inverseTransformation}
\end{align}
These transformations allow us to re-write the dynamical system in the following format
\begin{align}
	\begin{bmatrix}
		\dfrac{\mathrm{d} \mathscr{U}}{\mathrm{d} \zeta}\\
		\dfrac{\mathrm{d} \mathscr{S}}{\mathrm{d} \zeta}\\
		\dfrac{\mathrm{d} \mathscr{W}}{\mathrm{d} \zeta}\\
	\end{bmatrix}
	=&
	\begin{bmatrix}
		0 & 0 & 0\\
		0 & 0 & 0\\
		0 & 0 & -c\\
	\end{bmatrix}
	\begin{bmatrix}
		\mathscr{U}\\
		\mathscr{S}\\
		\mathscr{W}\\
	\end{bmatrix} \notag \\
	&+\dfrac{1}{c}
	\begin{bmatrix}
		-\left[\mathscr{S}(\mathscr{U}+\mathscr{W})\right]\\[5pt]
		\left[\mathscr{S}(r_1c\mathscr{U}+r_2\mathscr{S})\right]\\[5pt]
		\left[(\mathscr{U}+\mathscr{W})\left[-r_1c\mathscr{U} + (1-r_2)\mathscr{S}\right] + c^2\mathscr{U}(1+c\mathscr{U})\right]\\[5pt]
	\end{bmatrix}. \label{eq:dUdSdWtransformed}
\end{align}
To find the slow manifold we take the usual approach of writing the fast dynamics associated with $\lambda_1$ as a function of the slow dynamics that are associated with the zero eigenvalues by assuming that slow manifold can be locally expressed as a quadratic in $\mathscr{U}$ and $\mathscr{V}$. Equating coefficients with the tangency condition~\cite{Wiggins2003} gives the slow manifold,
\begin{equation}
\mathscr{W}(\mathscr{U},\mathscr{S})=\dfrac{1}{c^2}\left[ c(c^2-r_1)  \mathscr{U}^2 + (1-r_2)  \mathscr{U}\mathscr{S}\right], \label{eq:slowManifoldWscriptinUSscript}
\end{equation}
and the dynamics on the slow manifold are given by
\begin{align}
	\dfrac{\mathrm{d} \mathscr{U}}{\mathrm{d} \zeta}&=-\dfrac{1}{c^3}\left[c(c^2-r_1)\mathscr{U}^2\mathscr{S}+(1-r_2)\mathscr{U}\mathscr{S}^2+c^2\mathscr{S}\mathscr{U}\right], \label{eq:ODEdUscriptslowManifold}\\
	\dfrac{\mathrm{d} \mathscr{S}}{\mathrm{d} \zeta}&=\dfrac{1}{c}\left[r_2\mathscr{S}^2+r_1c\mathscr{U}\mathscr{S}\right]. \label{eq:ODEdSscriptslowManifold}
\end{align}
We can now re--write the slow manifold and the dynamics on the slow manifold in the original, unrotated coordinate system, giving
\begin{equation}
W(U,S)=\dfrac{1}{c^3}\left[(c^2-r_1)U^2 - (1-r_2) U S - c^2 U\right], \label{eq:slowManifoldWinUS}
\end{equation}
and
\begin{align}
\dfrac{\mathrm{d} U}{\mathrm{d} \zeta}&=\dfrac{1}{c^3}\left[(c^2-r_1)SU^2 -  (1-r_2) U S^2 - c^2SU\right],\label{eq:ODEdUslowManifold}\\
\dfrac{\mathrm{d} S}{\mathrm{d} \zeta}&=\dfrac{1}{c} \left[r_2 S^2 - r_1 U S\right].\label{eq:ODEdSslowManifold}
\end{align}
With these tools we may now plot the phase space including the two equilibrium points, and superimpose the slow manifold and the heteroclinic orbit obtained be re-writing the long--time PDE solution in terms of the $(U(\zeta), S(\zeta), W(\zeta))$ coordinates.  This information is summarised in Figure \ref{fig:7} for two smooth--fronted travelling waves and one sharp--fronted travelling wave, each with $r_1=r_2=1$.  Before considering Figure \ref{fig:7} in detail, note that a small $S$ and $U$ analysis of (\ref{eq:ODEdUslowManifold})--(\ref{eq:ODEdSslowManifold}) shows that the heteroclinic orbit must have $U \sim (r_2+1)S/r_1$ as $S \to 0^+$, meaning that the slope of the heteroclinic orbit is $r_1/(r_2+1)$ in the $US$--plane near the origin, and $U \sim A \textrm{exp}(-z/c)$ and $S \sim B \textrm{exp}(-z/c)$, for some constants $A>0$, $B>0$, as $z \to \infty$ for smooth--fronted travelling wave solutions.  These results for the flow on the slow manifold confirm (\ref{eq:Ulargez})--(\ref{eq:Slargez}) with $c=1/b$.
Figure \ref{fig:7}(a) shows the three--dimensional desingularised phase space together with the invaded equilibrium point in green, the uninvaded equilibrium point in blue, the heteroclinic orbit in solid blue and the slow manifold in grey.  In this case we have $c_2 = 10$ and we see that, as expected, the heteroclinic orbit enters the uninvaded equilibrium point after moving along the slow manifold.  In Figure \ref{fig:7}(b) we plot the slow manifold locally around the uninvaded equilibrium point together with the vector field defined by Equations (\ref{eq:ODEdUslowManifold})--(\ref{eq:ODEdSslowManifold}).  The heteroclinic orbit from the long--time PDE solution is shown in blue.  We see that the heteroclinic orbit is tangential to the vector field and enters the uninvaded equilibrium point.  For completeness we also solve Equations (\ref{eq:ODEdUslowManifold})--(\ref{eq:ODEdSslowManifold}) numerically to show a number of other solution trajectories on the slow manifold in red.  While these other solution curves are valid solutions of Equations (\ref{eq:ODEdUslowManifold})--(\ref{eq:ODEdSslowManifold}), they are unphysical in the sense that they are not associated with the travelling wave solution since they do not form a heteroclinic orbit joining the invaded and uninvaded equilibrium points. Figure \ref{fig:7}(b)--(c) shows a similar set of results to those in Figure \ref{fig:7}(a)--(b) for a different smooth--fronted travelling wave, this time with $c_1 = 1$.  Again we see that the heteroclinic orbit moves into the uninvaded equilibrium point along the slow manifold in Figure \ref{fig:7}(c), with additional details shown on the slow manifold in Figure \ref{fig:7}(d).  Interestingly,  results in Figure \ref{fig:7}(e)--(f), for a sharp--fronted travelling wave with $c_{\rm{min}} = 0.29$ are quite different to the smooth--fronted travelling waves in Figure \ref{fig:7}(a)--(d).  Here the heteroclinic orbit joining the invaded and uninvaded equilibrium points enters the uninvaded equilibrium point directly, without moving along the slow manifold.  This difference is highlighted in Figure \ref{fig:7}(d) where we see that there is no component of the heteroclinic orbit on the slow manifold.  These results in Figure \ref{fig:7} are for one particular choice of $r_1=r_2=1$, and similar results for different choices of $r_1$ and $r_2$ show the same qualitative behaviour (Supplementary Material).

\begin{figure}[h!]
	\centering
	\includegraphics[width=0.85\linewidth]{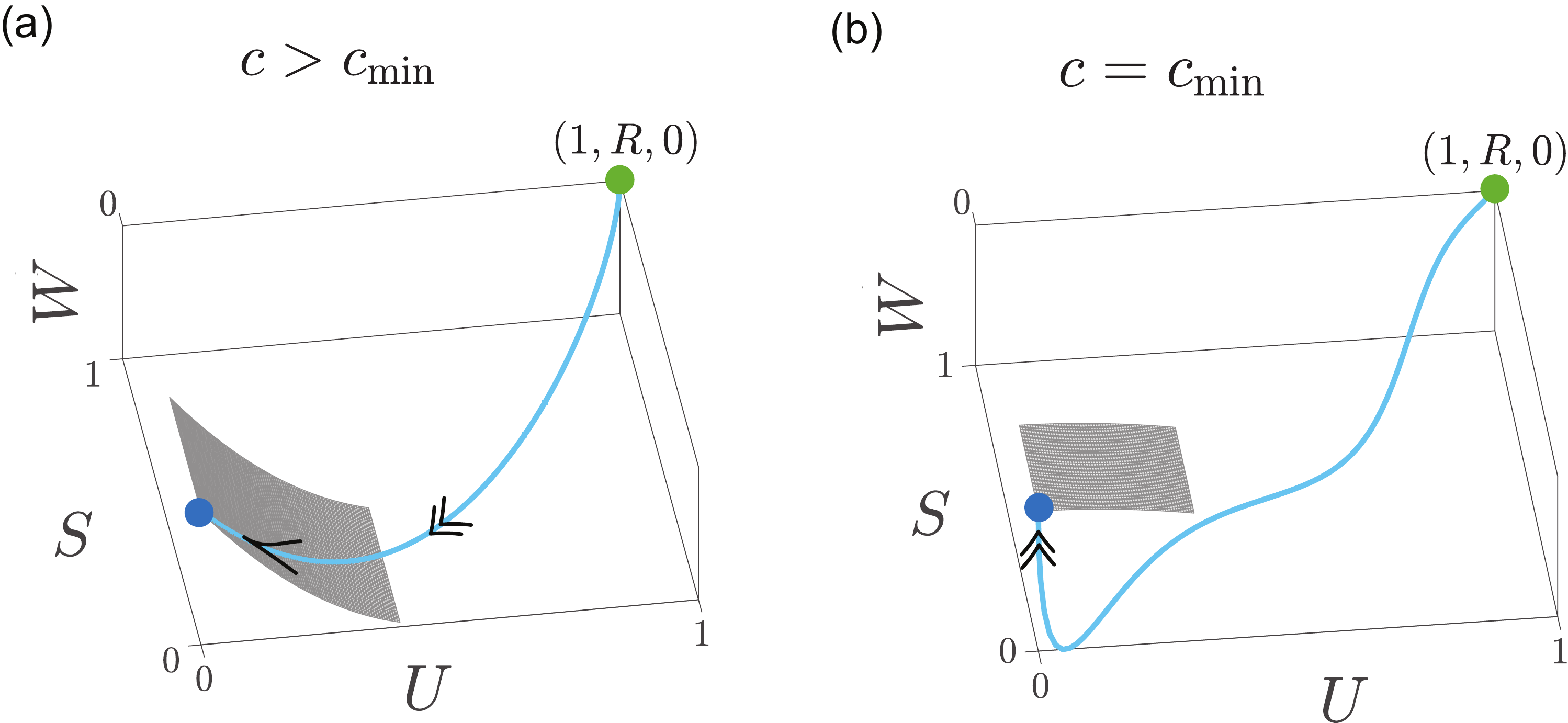}
	\caption{\textbf{Schematic distinction between smooth--fronted and sharp--fronted travelling wave solutions.} The schematic in (a) corresponds to a smooth--fronted travelling wave solution with $c > c_{\rm{min}}$, where the heteroclinic orbit (blue) in the desingularised phase space moves into the $(0,0,0)$ along the slow manifold (grey surface).  The schematic in (b) corresponds to a sharp--fronted travelling wave solution with $c = c_{\rm{min}}$, where the heteroclinic orbit (blue) enters the uninvaded equilibrium point, $(0,0,0)$, without moving along the slow manifold (grey surface).}	
	\label{fig:8}
\end{figure}

In summary, these results show us that we can make a simple geometric distinction between smooth--fronted travelling waves and sharp--fronted travelling waves using the slow manifold reduction.  Smooth--fronted travelling waves involve a heteroclinic orbit joining $(\bar{U},\bar{S},\bar{W}) = (1,R,0)$ and  $(\bar{U},\bar{S},\bar{W}) = (0,0,0)$, such that the heteroclinic orbit enters $(0,0,0)$ along the slow manifold, given by Equation (\ref{eq:slowManifoldWinUS}).  In contrast, sharp--fronted travelling waves involve a heteroclinic orbit joining the same two equilibrium points, with the difference being that the heteroclinic orbit enters $(0,0,0)$ directly, without moving along the slow manifold.  These differences are summarised schematically in Figure \ref{fig:8}.

It is worth noting that the computational phase space tools in Figure \ref{fig:7}(a),(c) and (e) provide physical insight into the interpretation of the minimum wave speed, $c_{\textrm{min}}$, for the substrate model.  While it is not possible to compute a long--time PDE solution with $c < c_{\textrm{min}}$, it is straightforward to plot the three--dimensional phase space and integrate Equations (\ref{eq:ODEdUdzeta})--(\ref{eq:ODEdWdzeta}) numerically to explore various trajectories in the relevant octant where $U \ge 0$, $S \ge 0$ and $W \le 0$.  These explorations show that we can identity a unique trajectory that enters the origin just like we did for $c \ge c_{\textrm{min}}$, however part of this trajectory has $U < 0$ which is why it can never be associated with a physically relevant travelling wave solutions (Supplementary Material). This observation shares similarities and differences with the phase plane analysis of the classical Fisher-KPP model, where the exact result $c_{\textrm{min}} = 2$ is found by ensuring that $U > 0$ near the origin~\cite{Murray2002}.  In the simpler Fisher-KPP model, the origin is an equilibrium point and so linearisation gives us the local properties of the phase plane, leading to this result.  Similar methodology applies for more complicated generalisations of the Fisher-KPP model~\cite{Vittadello2018}.   In the case of our substrate model, it appears that $c_{\textrm{min}}$ is also defined by requiring that $U > 0 $ along the heteroclinic orbit (Supplementary Material).  Conversely, numerical explorations show that when $c < c_{\textrm{min}}$ we observe that $U < 0$ for portions of the orbit that do not pass through a neighbourhood of the equilibrium point.  This observation suggests that linearisation cannot be used to find a mathematical expression for $c_{\textrm{min}}$.

For the next part of this work we attempt to understand how the shape of the travelling wave profiles depends upon the parameters in the mathematical model.  We will derive two such approximations; one for sharp--fronted travelling wave solutions, and the other for smooth--fronted travelling wave solutions.   In both cases we test our approximations using full time--dependent PDE solutions.

\subsection{Approximate solution for sharp--fronted travelling waves}\label{sec:approximationr1andr2big}
Numerical results in Section \ref{sec:1Dexploration} imply a relationship between the substrate model and the Porous--Fisher model, which we now explore further.  For fast substrate production and decay, $r_1\gg 1$ and $r_2\gg 1$, respectively, we anticipate that Equation (\ref{eq:GouvdiffSNonDimensional}) gives approximately $s = Ru$, and that Equation (\ref{eq:GouvdiffUNonDimensional}) is approximately
\begin{equation}
\label{eq:GouvdiffPorousFisherNonDimensional}
\dfrac{\partial u}{\partial t}= R\dfrac{\partial}{\partial x}\left(u \dfrac{\partial u}{\partial x}\right) + u(1-u), \quad 0 < x < \infty,
\end{equation}
which is the non--dimensional Porous--Fisher model with the diffusion term scaled by the constant $R$.  Therefore, we can make use of known results for the Porous--Fisher model in this limit.  In particular, sharp--fronted travelling wave solutions of the Porous--Fisher model are known to have the closed--form solution~\cite{Murray2002,Sherratt1996}
\begin{align}
U(z)&=
\begin{cases}
1-\textrm{exp}\left(\dfrac{z-z_c}{2c}\right), & z<z_c,\\
0, & z > z_c, \label{eq:exactPorousFisherUz}
\end{cases}\\
S(z) &= R U(z) \quad \quad \, \, \, -\infty < z < \infty, \label{eq:exactPorousFisherSz}
\end{align}
where $c = c_{\rm{min}} = \sqrt{R/2}$ and $z_c$ is the location of the sharp front~\cite{Murray2002}.   Note that Equation (\ref{eq:exactPorousFisherSz}) is equivalent to substituting Equation (\ref{eq:exactPorousFisherUz}) into Equation (\ref{eq:SzasfunctionIntU}) and evaluating the resulting expression in the limit that $r_1 \to \infty$ and $r_2 \to \infty$.

\begin{figure}[H]
	\centering
	\includegraphics[width=1\linewidth]{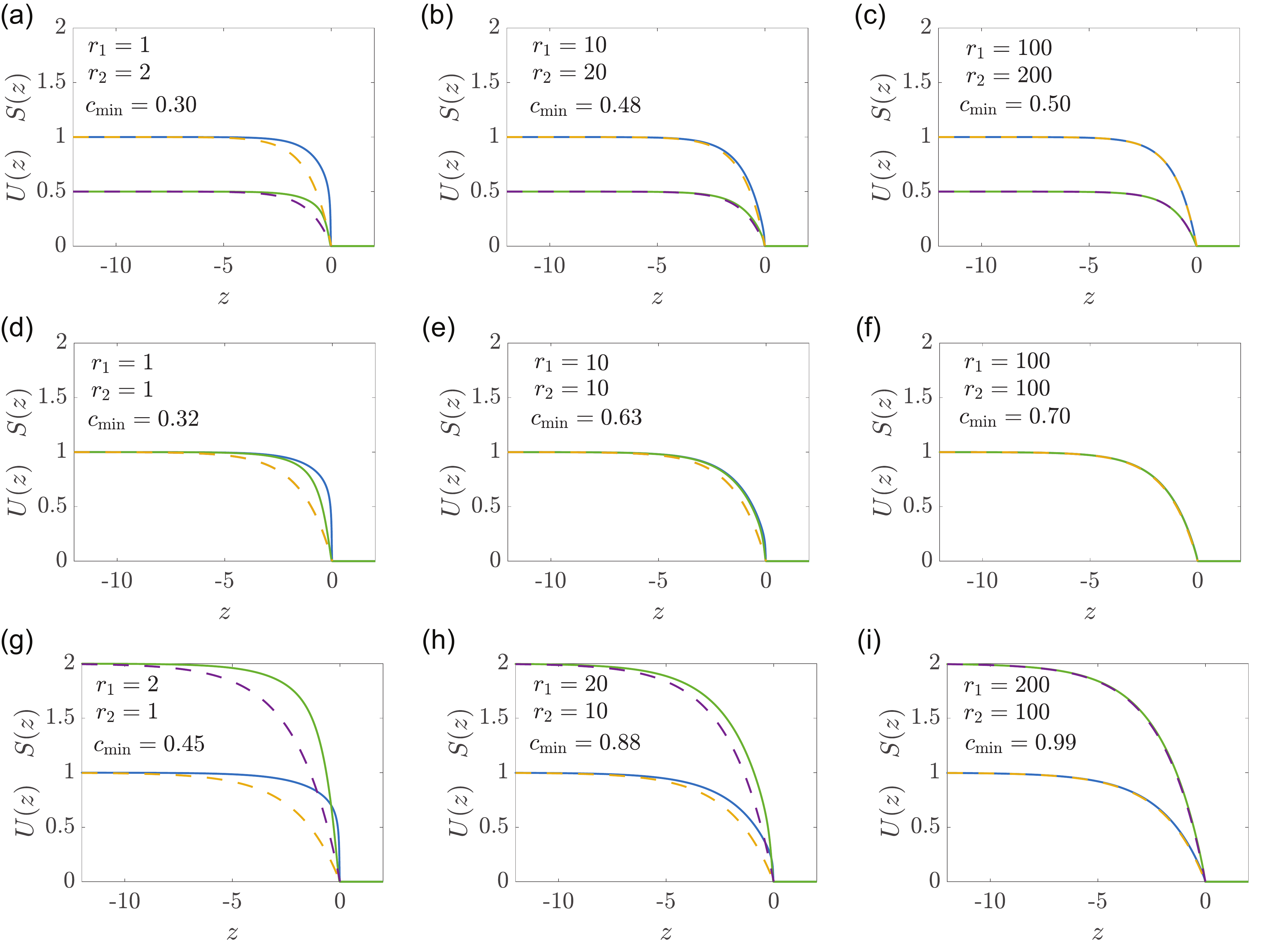}
	\caption{\textbf{Approximate shape of sharp--fronted travelling wave solutions.} Various numerical travelling wave solutions,  $U(z)$ and $S(z)$, obtained by solving Equations (\ref{eq:GouvdiffUNonDimensional})--(\ref{eq:ICSCompactSupport}) are compared with the approximation given by Equations (\ref{eq:exactPorousFisherUz})--(\ref{eq:exactPorousFisherSz}), where $z$ is shifted so that $z_c=0$.  Results in (a)--(c), (d)--(f) and (g)--(i) correspond to $R=0.5, 1$ and $2$, respectively.  Each subfigure shows the appropriate value of $r_1$, $r_2$ and $c_{\rm{min}}$. All numerical PDE solutions correspond to  $\Delta x = 1\times10^{-2}$, $\Delta t = 1\times10^{-3}$ and $\epsilon=1\times10^{-10}$.}
	\label{fig:9}
\end{figure}

Results in Figure \ref{fig:9} examine how late--time numerical PDE solutions can be approximated by Equations (\ref{eq:exactPorousFisherUz})--(\ref{eq:exactPorousFisherSz}).  Results in (a)--(c), (d)--(f) and (g)--(i) correspond to $R=0.5, 1$ and $2$, respectively, and in each case we see that Equations (\ref{eq:exactPorousFisherUz})--(\ref{eq:exactPorousFisherSz}) provide a good match with the shape of the travelling wave solution of the substrate model as $r_1$ and $r_2$ increase.

\subsection{Approximation solution for smooth--fronted travelling waves}\label{sec:approximationcbig}
Previous results in Figure \ref{fig:8} suggest that smooth--fronted travelling waves become less steep as $c$ increases, implying that $W(z)=\mathrm{d}U / \mathrm{d}z \to 0$ as $c \to \infty$.   Following the work of Canosa we make use of this observation to develop a perturbation solution by re-scaling the independent variable, $\hat{z} = z/c$ to give~\cite{Canosa1973},
\begin{align}
\dfrac{1}{c^2}\dfrac{\mathrm{d}}{\mathrm{d}\hat{z}}\left(S \dfrac{\mathrm{d}U}{\mathrm{d}\hat{z}}\right) + \dfrac{\mathrm{d}U}{\mathrm{d}\hat{z}} + U(1-U) &= 0,&-\infty < \hat{z} < \infty, \label{eq:ODEUcinfinity}\\
\dfrac{\mathrm{d} S}{\mathrm{d} \hat{z}} + r_1 U- r_2 S&= 0,&-\infty < \hat{z} < \infty.\label{eq:ODEScinfinity}
\end{align}
To proceed, we seek a perturbation solution in terms of the small parameter $1/c^2$ by expanding the dependent variables in a power series~\cite{Murray1984},
\begin{equation}
U(\hat{z}) = \sum_{n=0}^{\infty}c^{-2n} U_n(\hat{z}), \quad S(\hat{z}) = \sum_{n=0}^{\infty}c^{-2n} S_n(\hat{z}).
\end{equation}
Substituting these power series into Equations (\ref{eq:ODEUcinfinity})--(\ref{eq:ODEScinfinity}) and truncating after the first few terms gives
\begin{align}
&\dfrac{\mathrm{d}U_0}{\mathrm{d}\hat{z}} + U_0(1-U_0) = 0, \label{eq:ode1cinfinity}\\
&\dfrac{\mathrm{d} S_0}{\mathrm{d} \hat{z}} + r_1 U_0  - r_2 S_0  = 0 , \label{eq:ode2cinfinity}\\
&\dfrac{\mathrm{d} }{\mathrm{d}\hat{z}}\left(S_0\dfrac{\mathrm{d} U_0}{\mathrm{d}\hat{z}}\right) + \dfrac{\mathrm{d} U_1}{\mathrm{d}\hat{z}} + U_1(1-2U_0)= 0, \label{eq:ode3cinfinity}
\end{align}
with boundary conditions $U_0 \rightarrow 1$, $U_1 \rightarrow 0$ and $S_0 \rightarrow R$ as $\hat{z}\rightarrow -\infty$, and $U_0 \rightarrow 0$, $U_1 \rightarrow 0$ and $S_0 \rightarrow 0$ as $\hat{z}\rightarrow \infty$.  It is straightforward to solve these differential equations for $U_0(\hat{z})$, $U_1(\hat{z})$ and $S_0(\hat{z})$, however additional terms in the perturbation solution are governed by differential equations that do not have closed--form solutions.  Regardless, as we shall now show, these first few terms in the perturbation solution provide accurate approximations, even for relatively small values of $c$.

The solution of Equation (\ref{eq:ode1cinfinity}) is
\begin{align}
U_0(z) = \dfrac{1}{1+\textrm{exp}\left(\hat{z}\right)},
\end{align}
where we have arbitrarily chosen the integration constant so that $U_0(0)=1/2$. Given $U_0(z)$, we solve (\ref{eq:ode2cinfinity}) using an integrating factor to give
\begin{equation}
S_0(\hat{z}) = -r_1 \textrm{exp}\left(r_2 \hat{z}\right)\int_{\hat{z}}^{\infty}\dfrac{\textrm{exp}\left(-r_2 \hat{z}\right)}{1 + \textrm{exp}\left(\hat{z}\right)} \, \mathrm{d}\hat{z}. \label{eq:S0exprintegral}
\end{equation}
If $r_2$ is an integer we obtain
\begin{equation}
S_0(\hat{z}) = (-1)^{r_2}\textrm{exp}\left(r_2 \hat{z}\right)r_1\left[\ln\left(\textrm{exp}\left[-\hat{z}\right]+1\right)+\sum_{n=1}^{r_2}\dfrac{\textrm{exp}\left( -n \hat{z}\right)}{n\left(-1\right)^n}\right].
\end{equation}
If $r_2$ is not an integer there is no closed--form expression for $S_0(\hat{z})$ that we could find.  For particular integer choices of $r_1$ the expression for $S_0(\hat{z})$ is quite simple.  For example, with $r_2=1$ we have $S_0(\hat{z}) = r_1\left[1-\textrm{exp}\left(\hat{z}\right)\ln(\textrm{exp}\left[-\hat{z}\right]+1)\right]$, whereas for $r_2=2$ we have  $S_0(\hat{z}) =r_1\left[1/2-\textrm{exp}\left(\hat{z}\right)+\textrm{exp}\left(2\hat{z}\right)\ln(\textrm{exp}\left[-\hat{z}\right]+1)\right]$.
The solution for $U_1(\hat{z})$ is obtained by integrating Equation (\ref{eq:ode3cinfinity}) using an integrating factor to give
\begin{equation}
U_1(\hat{z}) =  \dfrac{\textrm{exp}\left(\hat{z}\right)}{(1 + \textrm{exp}\left[\hat{z}\right])^2}\int_{\hat{z}}^{\infty} \dfrac{\mathrm{d} }{\mathrm{d}\hat{z}}\left[S_0\dfrac{\textrm{exp}\left(\hat{z}\right)}{(1+\textrm{exp}\left[\hat{z}\right])^2}\right]\left[\dfrac{(1 + \textrm{exp}\left[\hat{z}\right])^2 }{ \textrm{exp}\left(\hat{z}\right)}\right] \mathrm{d}\hat{z}. \label{eq:U1exprintegral}
\end{equation}
Since this expression for $U_1(\hat{z})$ depends upon the expression for $S_0(\hat{z})$, we can only obtain closed--form expressions for $U_1(\hat{z})$ for integer values of $r_2$.  In these cases, expressions for $U_1(\hat{z})$  are relatively complicated and so we include these expressions in the Supplementary Material.  We note that care is required when evaluating $U_1(\hat{z})$ since the expression is indeterminate for large $\hat{z}$.  We address this simply by expanding $U_1(\hat{z})$ in a Taylor series as $\hat{z} \to \infty$ and plotting the series expansion for large $\hat{z}$.

\begin{figure}[H]
	\centering
	\includegraphics[width=1\linewidth]{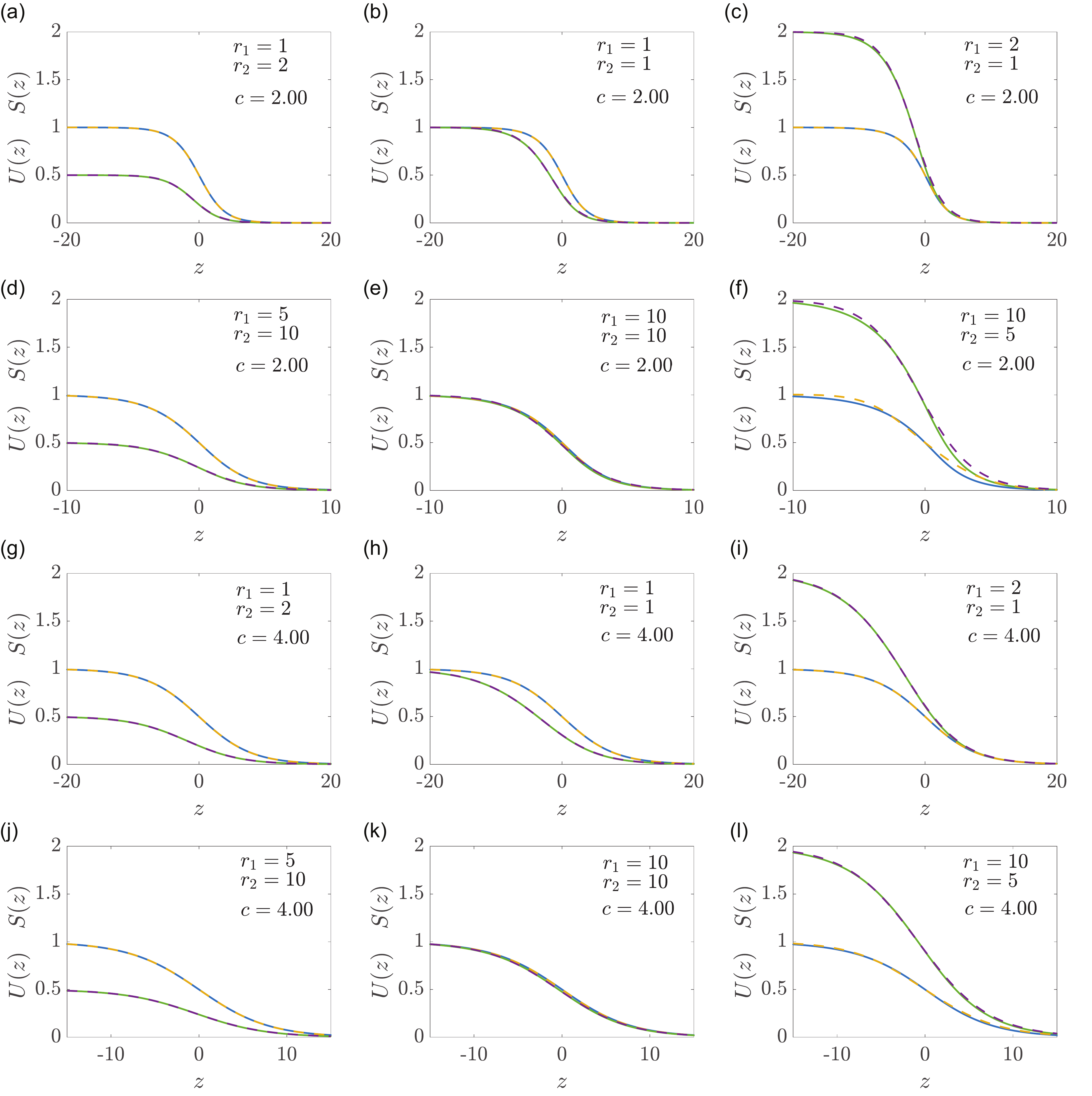}
	\caption{\textbf{Approximate shape of smooth--fronted travelling wave solutions.}  Results in (a)--(f) and (g)--(l) compare the numerical and perturbation solutions for $c=2.00$ and $c=4.00$, respectively.  Results in the left--most column correspond to $R=0.5$, those in the central column correspond to $R=1$, and those in the right--most column correspond to $R=2$.  Numerical solutions correspond to late--time numerical solutions of
Equations (\ref{eq:GouvdiffUNonDimensional})--(\ref{eq:BCNonDimensional}), with initial conditions given by Equations (\ref{eq:ICUNonCompactSupport})--(\ref{eq:ICSNonCompactSupport}) with appropriate values of $a$.  Numerical solutions of $U(z)$ and $S(z)$ are shown in blue and green, respectively.  Each numerical solution is superimposed with an $\mathcal{O}(1)$ perturbation solution for $S(z)$ and an $\mathcal{O}(1/c^2)$ for $U(z)$, and these perturbation solutions are shown in yellow and purple dashed curves, respectively.}	
	\label{fig:10}
\end{figure}

Results in Figure \ref{fig:10} compare the shapes of various smooth--fronted travelling wave solutions, for $c=2$ and $c=4$, with the $\mathcal{O}(1)$ perturbation solution for $S(z)$ and the $\mathcal{O}(c^{-2})$ perturbation solution for $U(z)$. These comparisons are made across a range of values of $r_1$ and $r_2$, and for $c=4$ the perturbation solutions are indistinguishable from the late--time numerical solutions.  In cases where $c=2$ we begin to see a small departure between the numerical and perturbation approximations.  Given that the perturbation solutions are valid in the limit $c \to \infty$, the quality of match in Figure \ref{fig:10} for $c=2$ and $c=4$ is quite good.

\newpage

\section{Conclusion and Future Work}
In this study we investigate a minimal model of cell  invasion that couples cell migration, cell proliferation and cell substrate production and decay.  A key feature of the mathematical model is that the diffusive flux is proportional to the substrate density so that the flux vanishes when the substrate is absent.  This feature leads to predictions of tissue formation involving the propagation of well--defined sharp fronts, and two--dimensional numerical simulations of the mathematical model recapitulate key features of recent experiments that involved the formation of thin tissues grown on D--printed scaffolds~\cite{Lanaro2021}.  To gain a deeper understanding of how the rate of substrate production and decay affects the rate of tissue production, the focus of this work is to study solutions of the substrate model in a one--dimensional geometry.

Preliminary numerical simulations of the substrate model in one dimension indicate that the mathematical model supports two types of travelling wave solutions.  As we show, sharp--fronted travelling waves that propagate with a minimum wave speed, $c_{\rm{min}}$, evolve from initial conditions with compact support, whereas smooth--fronted travelling waves that move with a faster wave speeds, $c > c_{\rm{min}}$, evolve from initial conditions where the density decays exponentially with position.  These numerical features are reminiscent of established features of travelling wave solutions of the well--known Porous--Fisher model.

Much of our analysis focuses on exploring the relationships between smooth--fronted and sharp--fronted travelling wave solutions, and here key features of the analysis of the substrate model are very different to the analysis of the Porous--Fisher model.  For example, there are three equilibrium points in the desingularised phase plane for the Porous--Fisher model whereby travelling wave solutions are characterised by a heteroclinic orbit that enters $(\bar{U},\bar{V}) = (0,0)$, whereas sharp--fronted travelling wave solutions involves a heteroclinic orbit that enters $(\bar{U},\bar{V}) = (0,-c)$.  In contrast, the desingularised phase space for the substrate model involves two equilibrium points only.  This means that both smooth--fronted and sharp--fronted travelling waves correspond to heteroclinic orbits that enter $(\bar{U}, \bar{S}, \bar{W}) = (0,0,0)$, which is fundamentally different to the Porous--Fisher model.  We provide a geometric interpretation that explains the difference between sharp--fronted and smooth--fronted travelling wave solutions since smooth--fronted travelling wave solutions are associated with a heteroclinic orbit that enters the origin in the desingularised phase space by moving along a slow manifold.  In contrast, sharp--fronted travelling wave solutions are associated with a heteroclinic orbit that enters the origin of the desingularised phase space directly, without moving along the slow manifold.  Additionally, we also develop and test useful closed--form expressions that describe the shape of the travelling wave solutions in various limits.  In particular, we provide accurate approximations for the shape of sharp--fronted travelling waves for sufficiently large $r_1$ and $r_2$, as well as accurate approximation of the shape of the smooth--fronted travelling wave solutions relevant for large $c$.

There are many avenues for extending the current work, and these options include further analysis of the current model as well as conducting parallel analysis for related mathematical models.  In terms of the current model, our analysis has not provided any relationship between $c_{\rm{min}}$ and the two parameters in the nondimensional model, $r_1$ and $r_2$. For simpler mathematical models, such as the Fisher-KPP model, the relationship between the minimum wave speed and the parameters in the model arises by linearising about the leading edge of the travelling wave~\cite{Murray2002}.  As we have shown, an interesting feature of the substrate model is that standard techniques to linearise about the leading edge do not apply.  Another possibility for extending the analysis of this model would be to consider the mathematical model in two-dimensions, such as describing the late--time dynamics of hole--closing phenomena~\cite{McCue2019}.

A different class of extensions of this work would be to consider generalising the nonlinear diffusion term in the substrate model, such as
\begin{align}
\label{eq:GouvdiffUNonDimensional2}
&\dfrac{\partial u}{\partial t}= \dfrac{\partial}{\partial x}\left(\mathcal{D}(s) \dfrac{\partial u}{\partial x}\right) + u(1-u), &0 < x < \infty\\
&\dfrac{\partial s}{\partial t}= r_1 u - r_2 s, &0 < x < \infty. \label{eq:GouvdiffSNonDimensional2}
\end{align}
This generalised substrate model involves a nonlinear diffusivity function, $\mathcal{D}(s)$.  We anticipate that nonlinear diffusivity functions with the property $\mathcal{D}(0) = 0$ will support sharp--fronted travelling wave solutions, and there are many such candidate functions.  One option of interest is a power--law diffusivity  $\mathcal{D}(s) = s^n$, where $n$ is some exponent.  It would be interesting to explore how different choices of $n$ affect various qualitative and quantitative features of the travelling wave solutions that have been established in the present study for $n=1$.  We hope to return to address these open questions in future research.

\noindent
\paragraph{Acknowledgements} This work is supported by the Australian Research Council (DP200100177). We thank Dr Oliver Maclaren for assistance with the slow manifold reduction, and we thank Dr Pascal Buenzli for helpful discussions about tissue engineering applications.

\noindent
\paragraph{Contributions} All authors conceived and designed the study and performed the mathematical analysis; M.El-H. performed numerical and symbolic calculations.  All authors drafted the article and gave final approval for publication.

\noindent
\paragraph{Competing Interests} We have no competing interests.

\end{spacing}

\end{document}